%% file: main.tex
\documentclass[letterpaper,twocolumn,10pt,dvipsnames]{article}
\input{pkg}

\usepackage{usenix-2020-09}
\usepackage[square,comma,numbers,sort&compress]{natbib}

\newcommand{\etal}{\textit{et al.}}
\usepackage{xstring}
\newcommand{\PP}[1]{
\vspace{2px}
\noindent{\bf \IfEndWith{#1}{.}{#1}{\IfEndWith{#1}{?}{#1}{#1.}}}
}

\begin{document}

\thispagestyle{fancy}
\fancyhead{}
\fancyhead[L]{In Proceedings of the 33rd USENIX Security Symposium, Philadelphia, PA, USA, Aug. 2024}

\input{hdr}
\input{abstract}
\input{intro}
\input{bkg}
\input{qual-method}
\input{quan-method}
\input{qual-results}
\input{quan-results}
\input{takeaways}
\input{discuss}
\input{relwk}
\input{conclusion}

\bibliographystyle{abbrvnat}
\footnotesize
\bibliography{ref,conf}

\vspace{50mm}
\input{append}

\end{document}

%% file: pkg.tex
\usepackage{graphicx}
\usepackage{xcolor}
\usepackage{tikz}
\usepackage{multirow}
\usepackage{setspace}
\usepackage{boldline}
\usepackage{subcaption}
\usepackage{adjustbox}
\usepackage{soul}
\usepackage{color}
\usepackage{threeparttable}
\usepackage{hyperref}
\usepackage{tcolorbox}
\usepackage{colortbl}
\usepackage{enumitem}
\usepackage{balance}
\usepackage{authblk}
\usepackage{fancyhdr}

\makeatletter
\def\UrlOrds{\do\=\do\/\do\\\do\.\do\:\do\*\do\-\do\_\do\~\do\'\do\"\do\%}
\def\UrlAlphabet{%
      \do\a\do\b\do\c\do\d\do\e\do\f\do\g\do\h\do\i\do\j%
      \do\k\do\l\do\m\do\n\do\o\do\p\do\q\do\r\do\s\do\t%
      \do\u\do\v\do\w\do\x\do\y\do\z\do\A\do\B\do\C\do\D%
      \do\E\do\F\do\G\do\H\do\I\do\J\do\K\do\L\do\M\do\N%
      \do\O\do\P\do\Q\do\R\do\S\do\T\do\U\do\V\do\W\do\X%
      \do\Y\do\Z}
\def\UrlDigits{\do\1\do\2\do\3\do\4\do\5\do\6\do\7\do\8\do\9\do\0}
\g@addto@macro{\UrlBreaks}{\UrlOrds}
\g@addto@macro{\UrlBreaks}{\UrlAlphabet}
\g@addto@macro{\UrlBreaks}{\UrlDigits}
\makeatother

\def\Snospace~{\S{}}

%% file: hdr.tex
\title{\Large \bf I Experienced More than 10 DeFi Scams: On DeFi Users' Perception of \\Security Breaches and Countermeasures
}

\author[1]{Mingyi Liu}
\author[2]{Jun Ho Huh}
\author[1]{HyungSeok Han}
\author[1]{Jaehyuk Lee}
\author[1]{Jihae Ahn}
\author[1]{\\Frank Li}
\author[3]{Hyoungshick Kim}
\author[1]{Taesoo Kim}

\affil[1]{\textit{Georgia Institute of Technology}}
\affil[2]{\textit{Samsung Research}}
\affil[3]{\textit{Sungkyunkwan University}}

\maketitle

%% file: abstract.tex
\begin{abstract}
Decentralized Finance (DeFi) offers a whole new investment experience and has quickly emerged as an enticing alternative to Centralized Finance (CeFi).
Rapidly growing market size and active users, however, have also made DeFi a lucrative target for scams and hacks, with 1.95 billion USD lost in 2023. Unfortunately, no prior research thoroughly investigates DeFi users' security risk awareness levels and the adequacy of their risk mitigation strategies.

Based on a semi-structured interview study ($N=14$) and a follow-up survey ($N=493$), this paper investigates DeFi users' security perceptions and commonly adopted practices, and how those affected by previous scams or hacks (DeFi victims) respond and try to recover their losses. Our analysis shows that users often prefer DeFi over CeFi due to their decentralized nature and strong profitability. Despite being aware that DeFi, compared to CeFi, is prone to more severe attacks, users are willing to take those risks to explore new investment opportunities.
Worryingly, most victims do not learn from previous experiences; unlike victims studied through traditional systems, DeFi victims tend to find new services, without revising their security practices, to recover their losses quickly. The abundance of various DeFi services and opportunities allows victims to continuously explore new financial opportunities, and this reality seems to cloud their security priorities. Indeed, our results indicate that DeFi users' strong financial motivations outweigh their security concerns -- much like those who are addicted to gambling. Our observations about victims' post-incident behaviors suggest that stronger control in the form of industry regulations would be necessary to protect DeFi users from future breaches.

\end{abstract}

%% file: intro.tex
\section{Introduction}
\label{s:intro}

Decentralized Finance (DeFi) is a financial ecosystem built on blockchain platforms like Ethereum. It provides a variety of financial services, mainly executed through smart contracts~\cite{EVM}, such as Decentralized Exchanges (DEXs) and lending services. This emerging field has attracted substantial interest, with its Total Value Locked (TVL) reaching a record high of 180 billion USD in November 2021~\cite{defillama}.

With its growing popularity, 
DeFi has become an attractive target for hacks and scams, which we also refer to as DeFi incidents.
The total loss resulting from DeFi incidents in 2023 exceeded 1.95 billion USD~\cite{de-fi-report}.
Popular DEXs such as Balancer, Curve Finance, and dYdX were breached due to smart contract vulnerabilities~\cite{balancer-hack-2023, curve-hack-2023, dydx-hack-2023}.
Additionally, over 263 rug-pull scams were reported in 2023~\cite{de-fi-report}.
The infamous LUNA meltdown (reported in 2022) failed to protect its stablecoin values, incurring a total loss of 60 billion USD~\cite{luna-meltdown}.

Despite reports being published about numerous DeFi incidents and their losses, people are still heavily using DeFi, and its TVL remains at 130 billion USD in 2024~\cite{defillama}. Such trends motivated this work and two intriguing aspects:
(1) despite the prevalence of security breaches being reported, why do people continue to use DeFi services,
and (2) whether people make decisions based on an adequate understanding of security risks and mitigation practices.
To date, there is no prior research that thoroughly investigates DeFi users' perceptions of security risks.
Prior efforts often focused on smart contract security~\cite{he:ccs:2019, wustholz:fse:2020, nguyen:icse:2020, choi:icse:2021}. 
Wang~\etal~\cite{wang:chi:2022} studied user perceptions of DeFi incidents. However, their investigations were limited to the scope of sandwich attacks.

To bridge this gap, we conducted a semi-structured interview with 14 DeFi users, including real-world victims, and a follow-up online survey ($N=493$), investigating DeFi users' security risk awareness levels and the adequacy of commonly employed security practices, and how real-world victims respond to DeFi breaches and mitigate risks. These studies have been designed to address the following four research questions:

\PP{RQ1: Why do people continue to use DeFi despite numerous DeFi incidents being reported?}
Traditional Centralized Finance (CeFi) or bank users typically abandon their banks if serious incidents are reported~\cite{silicon}. Our curiosity lies in understanding why DeFi users behave differently.

\PP{RQ2: What DeFi risks are users concerned about, and how do they mitigate these risks?}
DeFi systems introduce unique security risks~\cite{zhou:oakland:2023}. Our objective is to understand whether DeFi users have sufficient knowledge about the overall threat landscape, and the security controls that need to be used.

\PP{RQ3: How do victims respond to DeFi incidents?}
Another objective is to understand what actions are taken by victims after experiencing scams or hacks, and investigate the adequacy of their actions in recovering their losses. Our investigation extends to understanding how victims' perceptions change after the incident.

\PP{RQ4: How do DeFi users perceive regulation?}
Regulatory controls may be considered in the future to better protect DeFi users. To that end, we were driven to understand DeFi users' perceived benefits and concerns with respect to introducing regulatory controls.

Our qualitative and quantitative studies reveal that people choose DeFi services because they appreciate its decentralized nature, and the unique investment opportunities they offer (\textbf{RQ1}).
Many participants consider DeFi to be less secure than CeFi. However, such participants still appreciate the transparency and trustworthiness guarantees of DeFi services, and continue using them anyway. 
Participants are mainly concerned about rug-pull scams, volatility of crypto prices, and smart contract exploitation (\textbf{RQ2}). However, many of them do not know how to effectively mitigate such threats. 
They often believe that traditional security systems are applicable and equally effective in protecting them from DeFi threats. For instance, many participants falsely believe two-factor authentication (2FA) will mitigate rug-pull or smart contract exploitation types of breaches. 
Worryingly, many victims simply find other DeFi services, and continue to use them to recover their losses (\textbf{RQ3}). Such victims fail to take any other remedy actions, and do not revise their security practices. Surprisingly, more than half of the studied victims explained that their security perceptions did not change after the incident. It seems like their overwhelming financial motivations cloud their security priorities: many participants fall victim to multiple scam or hack incidents, and still use DeFi without carefully considering the security risks. 
As expected, more participants oppose the idea of regulating DeFi services (\textbf{RQ4}) -- mainly due to their concerns about paying additional tax and regulatory controls possibly jeopardizing the decentralization benefits of DeFi. Considering that paying tax is a civil obligation, however, and users' tendency to ignore security risks in pursuit of financial gains, our recommendation is to explore decentralized regulations and mandate strong security controls and practices to better protect DeFi users.

Taken together, these key findings represent the major contributions of the paper: a first formal study investigating DeFi users' security concerns and risk mitigation strategies, a list of security misconceptions that need to be addressed, and a thorough report on the security behaviors of DeFi victims and how they can be better protected in the future. 

%% file: bkg.tex
\section{Background} \label{s:bkg}
In this section, we begin by reviewing DeFi with its characteristics and financial services. We then introduce the hacks and scams that have occurred in the DeFi realm.

\subsection{Decentralized Finance (DeFi)}
DeFi is a financial ecosystem backed by a blockchain such as Ethereum.
These blockchains empower DeFi to have unique characteristics compared to Centralized Finance (CeFi), giving rise to the emergence of distinct DeFi services.

\PP{DeFi characteristic}
Most DeFi services are implemented as smart contracts~\cite{EVM} on top of a blockchain.
DeFi developers encode their protocols and functionalities into smart contracts and deploy them on the blockchain by sending smart contract creation transactions (txs).
DeFi users then interact with DeFi services by sending txs to smart contracts of DeFi services.
This makes DeFi services inherit characteristics of blockchain such as \emph{decentralization}, \emph{transparency}, and \emph{accessibility}.

In particular, DeFi is \emph{decentralized} due to its foundation on blockchain, which operates in a decentralized manner.
Furthermore, the use of smart contracts as the backbone of DeFi implementations,
along with recording of all tx histories on the blockchain, ensures the \emph{transparency} of DeFi.
Lastly, DeFi employs blockchain accounts, which anyone can create with a private key. 
DeFi operates through txs, and any account owner can initiate txs.
These enhance the \emph{accessibility} of DeFi.

\PP{DeFi service}
DeFi offers many financial services inspired by CeFi while demonstrating the DeFi characteristics.
Similar to CeFi, DeFi provides financial services such as exchanges and lending services.
However, DeFi services operate with distinct mechanisms.
For example, DEXs often utilize an automated market maker (AMM) mechanism~\cite{xu:csur:2023},
which automatically determines the exchange ratio based on the quantities of tokens in liquidity pools.
Additionally, DeFi lending services introduce a unique feature called flashloan~\cite{flash-loan}.
By leveraging the atomicity of blockchain transactions and the programmability of smart contracts,
a flashloan allows borrowers to obtain a loan only if they repay it along with its interests within a single transaction.
Notably, DeFi offers other various services such as NFTs, insurance, governances, stablecoins, and cross-chain bridges.

\subsection{DeFi Hacks}
Although the blockchain, the basis of DeFi, is secure, DeFi services might be insecure because of vulnerabilities in their implementations.
In this paper, we categorize DeFi hacks into four categories based on the exploited components in DeFi.

\PP{Smart contract exploit}
Due to the fact that most DeFi protocols are implemented through smart contracts,
numerous DeFi hacks have happened by smart contract exploits, resulting in at least 1.57 billion USD losses until May 1, 2022~\cite{zhang:icse:2023}.
For example, the DAO attack~\cite{DAO-attack} exploited a \emph{re-entrancy} vulnerability, which caused inconsistent state updates.
Furthermore, attackers have actively exploited \emph{price oracle manipulation} vulnerabilities, resulting from developers misusing price oracle APIs to get token prices in DeFi services.

\PP{Cross-chain bridge exploit}
To connect DeFi services operating on different blockchains,
DeFi employs \textit{cross-chain bridges},
which facilitate the exchange of assets between two blockchains.
Unfortunately, some attackers have identified vulnerabilities within these bridges and exploited them to
manipulate tokens without providing assets on the other blockchain.
For instance, the Wormhole bridge was exploited by this vulnerability and lost 320 million USD~\cite{wormhole-attack}.

\PP{Private key leakage}
Some DeFi hacks occurred due to private key leaks, as these private keys serve as passwords of DeFi accounts.
For example, Ronin network's private keys were stolen, resulting in 625 million USD losses~\cite{ronin-network-attack}.

\PP{DeFi front-end attack}
DeFi services are typically based on smart contracts and DeFi users should send txs on blockchain to interact with them.
This might be a big hurdle for regular DeFi users, and DeFi developers provide some web pages (front-ends) to improve their usability.
However, some attackers exploited the web pages to make users interact with the web pages to send tokens to attackers rather than trading with DeFi services~\cite{balancer-attack}.

\subsection{DeFi Scams}
Similar to CeFi users, DeFi users are also susceptible to various scams, including phishing. However, DeFi scams differ from CeFi scams due to the unique characteristics of DeFi.

\PP{Rug-pull}
In DeFi ecosystems, anyone, including scammers, can launch their own DeFi services.
Therefore, scammers create fraudulent services and persuade DeFi users to invest their money in these scams.
In the end, scammers abandon their projects and disappear with the funds from DeFi users.
We call such scams \textit{rug-pulls}.
In particular, there have been numerous rug-pulls involving DeFi scam tokens, resulting in losses exceeding 240 million USD~\cite{cernera:sec:2023}.

\PP{Stablecoin meltdown}
To connect DeFi with CeFi, DeFi developers introduced \textit{stablecoins}~\cite{klages:aft:2020}, which are pegged to fiat currencies
-- stablecoins such as USDT and USDC are pegged to the USD.
DeFi users believe that stablecoins are backed by an adequate reserve of fiat currencies or certain algorithms to uphold their values.
However, some stablecoins failed to maintain their values, leading DeFi users to panic sell significant amounts of these stablecoins,
ultimately resulting in a stablecoin meltdown.
For instance, the ``LUNA meltdown'' failed to keep the value of its stablecoin, UST, which was backed by an algorithm,
resulting in 60 billion USD losses~\cite{luna-meltdown}.

\PP{Phishing}
DeFi ecosystems are not immune to phishing attacks, similar to the CeFi ecosystems.
DeFi phishers often reach out to potential victims via email or social media,
attempting to obtain the victims' private keys or tricking them into initiating specific txs.
Specifically, DeFi phishers may deceive victims into sending txs that include hidden token approval txs,
thereby granting the phishers access to drain tokens from the victims' wallets.

\PP{Airdrop scam}
One particular phishing method in DeFi involves using airdrops, where developers distribute tokens to DeFi users to advertise their services.
Airdrop scammers exploit this process by sending their tokens to wallets of potential victims.
These potential victims, upon receiving these unexpected tokens, may visit phishing websites or DeFi services to investigate the activities happening in their wallets.
While they interact with phishing websites or DeFi services, they may leak their private keys or initiate some fraudulent txs as mentioned.
Therefore, they are more likely to fall victim than to traditional phishing via email or social media.

%% file: qual-method.tex
\section{Methodology} \label{s:method}
To investigate the motivations and risk perceptions of DeFi users relevant to our research questions, we conducted a two-phase study. Initially, we performed in-depth interviews with 14 users, obtaining qualitative insights into their experiences. Based on the findings from the first study, we executed a quantitative survey with 493 DeFi users to validate and expand our understanding. Ensuring ethical and responsible research practices, both study designs received thorough review and approval from our Institutional Review Board (IRB). Minimizing data collection, we only gathered necessary personal information and stored responses under pseudonyms to ensure anonymity. Importantly, participants were informed of their right to withdraw at any time.
Supplementary study materials, including the interview guide, codebook, and survey questionnaire, are available in a GitHub repository\footnote{\url{https://github.com/mingyiliu95/defi-user-study}}.

\subsection{Study 1: Semi-structured Interview}
We aimed to gain a comprehensive understanding of user perceptions of DeFi, including their views on hacks and scams. To achieve this objective, we conducted semi-structured interviews with DeFi users ($N=14$).

\PP{Design.}
The interview protocol was designed with four sections to sequentially address our research questions.
(1) For \textbf{RQ1}, we asked participants about their perceptions of DeFi.
Specifically, participants discussed their \textit{preference} between DeFi and CeFi, along with justifications for their choices.
(2) For \textbf{RQ2}, we explicitly inquired about participants' perceived DeFi risks and their mitigation strategies.
(3) For \textbf{RQ3}, participants who were victims shared their experiences and reactions to DeFi hacks or scams.
We also explored whether these incidents affected their perceptions of DeFi.
(4) For \textbf{RQ4}, we gathered participants' opinions on regulations that could mitigate DeFi hacks and scams.
Notably, to ensure the accuracy and coherence of our interviews, we conducted three pilot interviews and primarily revised wording for clarification based on their feedback.
We excluded these pilot interviews from our data analysis.

\PP{Recruitment.}
To ensure we interviewed actual DeFi users, we implemented a rigorous two-step recruitment process. Firstly, we advertised our study through popular channels within active DeFi communities like Twitter, Telegram, Discord, and through word-of-mouth. In our recruitment advertisement, we stated that the purpose of our study was to understand user perceptions of cryptocurrency trading. Secondly, potential participants completed a screening questionnaire designed to filter for active DeFi users over the age of 18. This questionnaire included questions about the decentralized application (dApp) usage, community involvement, and experiences with DeFi hacks or scams. Specifically, participants were asked:
a) if they had interacted with dApps;
if so, b) the names of the dApps they used most frequently;
c) their role in the DeFi community (e.g., regular users, dApp developers);
and d) if they had experienced DeFi hacks or scams.
We then invited participants who accurately listed dApps, including those who identified themselves as victims of DeFi incidents, for interviews.
Each participant received 20 USD as compensation.

\PP{Data collection and analysis.}
We conducted 14 online interviews via recorded video calls, averaging 65 minutes in length.
After transcribing the videos, two researchers independently coded each interview and discussed their codes to reach a consensus.
This coding process was iterated for all 14 interviews, resulting in 148 codes and a Cohen's $\kappa$ inter-coder reliability score~\cite{fleiss2013statistical} of 0.89.
As appended in~\autoref{a:code-saturation-results}, the interview study was deemed complete once code saturation was reached without new codes emerging that addressed the research questions.

\PP{Demographics.}
\autoref{t:interview-demo} presents the detailed demographics of our interview participants ($N=14$).
The sample was predominantly male and younger, but there was a varied representation in terms of income and educational background.
Participants ranged from newcomers to seasoned users in their cryptocurrency experience.
Additionally, we disclosed whether participants were DeFi developers, considering the potential for bias from them.
Lastly, we specifically targeted participants who had experienced DeFi misconduct, resulting in eleven self-identified victims in our interviews.

\begin{table}[t!]
    \begin{threeparttable}
        \centering
        \tabcolsep=1.8pt
        \footnotesize
        \caption{The demographics of interview participants.}
        \label{t:interview-demo}
        \begin{tabular}{l|ccccccc}
        \hlineB{3}
        \multirow{2}*{\textbf{ID}} & \multirow{2}*{\textbf{Gender}} & \multirow{2}*{\textbf{Age}} & \multirow{2}*{\textbf{Education}} & \multirow{2}*{\textbf{Income}} & \textbf{Crypto} & \multirow{2}*{\textbf{Dev}\tnote{2}} & \multirow{2}*{\textbf{Victim}} \\
                    & & & & & \textbf{YoE}\tnote{1} & & \\
        \hlineB{3}
        P1 & Female & 18-24 & Bachelor's & \$50k-75k & 3-5 & No & Yes \\
        \hline
        P2 & Female & 18-24 & Bachelor's & \$50k-75k & 1-3 & No & Yes \\
        \hline
        P3 & Male & 18-24 & Bachelor's & \$25k-50k & 1-3 & No & Yes \\
        \hline
        P4 & Male & 25-34 & After bachelor's & \$200k+ & 7-9 & Yes & Yes \\
        \hline
        P5 & Male & 25-34 & After bachelor's & \$50k-75k & 3-5 & No & Yes \\
        \hline
        P6 & Male & 25-34 & High school & $<$\$25k & 3-5 & No & No \\
        \hline
        P7 & Male & 25-34 & Bachelor's & $<$\$25k & 3-5 & No & Yes \\
        \hline
        P8 & Male & 25-34 & Bachelor's & \$25k-50k & 3-5 & No & Yes \\
        \hline
        P9 & Male & 35-44 & After bachelor's & \$100k-125k & 3-5 & No & Yes \\
        \hline
        P10 & Male & 18-24 & Bachelor's & \$150k-175k & 3-5 & No & Yes \\
        \hline
        P11 & Male & 18-24 & Bachelor's & N/A & 3-5 & Yes & No \\
        \hline
        P12 & Male & 18-24 & High school & $<$\$25k & 1-3 & Yes & Yes \\
        \hline
        P13 & Male & 18-24 & High school & N/A & 3-5 & No & No \\
        \hline
        P14 & Male & 25-34 & Bachelor's & \$75k-100k & 1-3 & No & Yes \\
        \hlineB{3}
        \end{tabular}
        \begin{tablenotes}
           \item $^1$ Abbreviated for \textit{Years of Experience}; $^2$ Abbreviated for \textit{Developer}.
        \end{tablenotes}
    \end{threeparttable}
\end{table}

%% file: quan-method.tex
\subsection{Study 2: Large-scale Survey}

Based on the results of the interview study, we conducted a quantitative study via an online survey ($N=493$) on Prolific\footnote{https://www.prolific.com/} to statistically validate our observations from the interviews on a large scale.

\PP{Design.}
We structured the survey into four sections, mirroring our interview design.
(1) For \textbf{RQ1}, we inquired about participants' positive and negative experiences with DeFi, and their \textit{preference} for DeFi over CeFi. Specifically, we assessed factors such as \textit{security}, \textit{usability}, \textit{transparency}, and \textit{trust}, identified in the interviews as influencing DeFi preferences.
(2) Addressing \textbf{RQ2}, we asked participants to select and rank the most concerning DeFi risks identified from interviews. We then queried about the countermeasures they had adopted to mitigate these risks.
(3) When tackling \textbf{RQ3}, we probed participants' unfortunate experiences with DeFi hacks or scams, including the type of incident, their remedial actions, and any changes in perception. We did not pre-select victims for our survey to accurately represent the real-world proportion of users affected by DeFi incidents.
(4) For \textbf{RQ4}, we gathered participants' views on DeFi regulations, asking them to express their support or opposition to regulatory oversight and explain their reasons. We included demographic questions at the survey's start and inserted two attention-checking questions within the sections on RQ1 and RQ3.

\PP{Recruitment.}
Because there was no pre-defined filter on Prolific to screen DeFi users, we used a screening questionnaire to recruit DeFi users for the full survey. Initially, we applied three built-in Prolific filters to ensure participants a) had used cryptocurrencies, b) maintained an approval rate above 95\% for past submissions, and c) were U.S. residents to minimize cultural and regulatory differences. In the screening questionnaire, we asked if participants considered themselves DeFi users and, for validation, to name the dApp they most frequently used along with its smart contract address. Respondents who accurately listed dApps were deemed eligible and invited to participate in the full survey. We added the Prolific IDs of eligible participants to an allowlist, ensuring only those selected could take the survey.
Despite a significant drop-off, this screening process was necessary to ensure the validity of the participants.
Compensation was set at 0.30 and 2.50 USD for completing the screening questionnaire and full survey, respectively.

\PP{Data collection and analysis.} From June to October 2023, we received 4,380 responses to the screening questionnaire. We invited 1,134 eligible participants, of whom 550 submitted the full survey, averaging a completion time of 15 minutes. We analyzed 493 valid responses, excluding incomplete submissions, failed attention checks, or non-compliant responses (e.g., out-of-range ranks).
Due to the non-normal distribution of the collected data, we employed non-parametric statistical tests for our analysis. We performed Mann-Whitney U tests and Chi-squared tests of independence (each at a significance level of $\alpha=0.05$) to compare two answers (questionnaire options). Since each answer was compared to every other answer pairwise, we applied Bonferroni correction. In addition to $p$-values, we computed rank-biserial correlation $r$\footnote{The thresholds for interpreting effect sizes as small, medium, and large are 0.10, 0.30, and 0.50, respectively.} to report the effect size.

\PP{Demographics.} The demographics of our survey participants ($N=493$) are presented in \autoref{t:survey-demo}. Our sample predominantly consisted of males (82.4\%) and individuals under 44 years old (78.0\%), with average and median ages of 37.75 and 35, respectively. Furthermore, 77.0\% held a degree at or above a Bachelor's level. Although we omitted specific occupation distribution due to diverse responses, ``Computer and Mathematical'' occupations were the most common (15.6\%).

To assess the representativeness of our participants, given the scarcity of quantitative user studies in the DeFi arena, we compared our demographics with those from research targeting general crypto-asset users~\cite{abramova:chi:2021}. Our study had a higher proportion of male participants than the reference (77.5\%) and a younger demographic, with the majority aged between 25 and 34 years (38.3\%), as opposed to the reference study's primary age group of 35-44 years (36.2\%). Our respondents also had higher educational attainment than the general U.S. population~\cite{us-facts}, aligning closely with the referenced study, where 77.2\% of participants had at least a Bachelor's degree.

\begin{table}[t!]
    \centering
    \footnotesize
    \caption{The demographics of survey participants.}
    \label{t:survey-demo}
    \begin{tabular}{llc}
    \hlineB{3}
    \multirow{2}*{\textbf{Item}} & \multirow{2}*{\textbf{Property}} & All ($N$=493) \\
                               &                   & \textbf{\% of participants} \\
    \hlineB{3}
    \multirow{4}*{Gender}      & Male              & 82.4 \\
                               & Female            & 16.0 \\
                               & Non-binary        & 1.2  \\
                               & No answer         & 0.4  \\
    \hline
    \multirow{6}*{Age}         & 18-24           & 9.7  \\
                               & 25-34           & 38.3 \\
                               & 35-44           & 30.0 \\
                               & 45-54           & 14.2 \\
                               & 55-64           & 6.1  \\
                               & 65 or above       & 1.6  \\
    \hline
    \multirow{7}*{Education}   & No schooling      & 0.0  \\
                               & No high school    & 0.0  \\
                               & High school       & 21.7 \\
                               & Bachelor's        & 59.2 \\
                               & After bachelor's  & 17.8 \\
                               & Other             & 0.8  \\
                               & No answer         & 0.4  \\
    \hline
    \multirow{10}*{Income}     & $<$\$25k   & 6.5  \\
                               & \$25k-50k       & 17.4 \\
                               & \$50k-75k       & 22.9 \\
                               & \$75k-100k      & 18.7 \\
                               & \$100k-125k     & 11.8 \\
                               & \$125k-150k     & 8.9  \\
                               & \$150k-175k     & 3.4  \\
                               & \$175k-200k     & 2.4  \\
                               & \$200k+   & 6.5  \\
                               & No answer & 1.4  \\
    \hlineB{3}
    \end{tabular}
\end{table}

%% file: qual-results.tex
\section{Results}
\label{s:results}

This section presents the study results addressing our research questions. We report qualitative findings in~\autoref{s:qual-results}, quantitative validations in~\autoref{s:quan-results}, and summarize key takeaways in~\autoref{s:takeaways}.

\subsection{Interview Study Results}
\label{s:qual-results}

\subsubsection{Perceptions of DeFi}
\label{s:qual-rq1}

Focusing on tackling \textbf{RQ1}, we compared users’ perceptions of DeFi and CeFi and analyzed why users prefer or do not prefer DeFi over existing CeFi services.

\PP{Preference}
Most interview participants preferred DeFi over CeFi for various reasons: a few participants highlighted ease of access and use as the main reason, while some cited additional features, transparency, and reliability as the reasons for preferring DeFi. For example, P14 highlighted accessibility benefits: ``\textit{I don't have to open up an account. I just need a unique wallet.}'' A small number of participants shared neutral or more negative feedback on DeFi, explaining their concerns related to security issues, excessive fees, and lack of experience. P2, for instance, explained ``\textit{Since I'm a beginner [in trading], I would go with traditional [fiat CeFi]}.''
We identify security, usability, transparency, trust, and profitability as common themes affecting preference to use DeFi, and provide example quotes below.

\PP{Security}
Only a few interview participants mentioned security as a reason for preferring DeFi: ``\textit{DeFi is a more secure way to make transactions}'' (P3).
Half of those who did not prefer DeFi emphasized its insecurity. P13 explained ``\textit{Because [of] the amount of hacks... it sounds like it would be very difficult to secure [DeFi].}''

\PP{Usability}
In response to the question about reasons for preferring to use DeFi, some interview participants mentioned that DeFi is easier to use, referring to fast transaction and interoperability advantages.
For example, P5 explained ``\textit{[DeFi is] easier to use... You can go to DeFi and actually make quick transactions.}'' P10 mentioned the ease in which crypto assets can be transferred: ``\textit{exchange from one protocol to another [is] easier compared to stock exchange.}'' Several participants also mentioned the convenience factor associated with not having to verify their identity to set up and activate accounts. Among those who preferred DeFi for reasons other than usability, some mentioned their concerns about the steep learning curve.

\PP{Transparency}
Half of those interview participants who preferred to use DeFi mentioned transparency. P4 explained ``\textit{No backroom dealings... Everything is public.}'' 
These participants expressed concerns about the opaqueness of the practices employed by crypto and fiat CeFi services: ``\textit{You don’t know what the bank’s security is [but] you know what the DeFi security is... it’s all on chain}'' (P11).

\PP{Trust}
Some preferred to use DeFi because they do not trust the operations of fiat CeFi services, ``\textit{[Fiat CeFi] is not nearly as safe as everybody thinks it is. I don't trust the regular financial system}'' (P11).
On the contrary,
one interview participant explained that he lost trust in DeFi and stopped using it: ``\textit{I don't use any of them [dApps] anymore... I don't really trust them apart from hacks}'' (P8). 

\PP{Profitability}
Many participants acknowledged the profitability side of DeFi as the main reason for preferring to use DeFi. P7, for example, explained financial opportunities: ``\textit{There are a lot of opportunities to have a twenty percent APY income, so I put my money to the DeFi and earn the yield.}'' However, there were also several participants who expressed expensive transaction fees as a reason for carefully choosing dApps: ``\textit{Fees are important. If I see competitive [tx] fees, I like to use them [more] often}'' (P6).

\begin{figure}[t!]
    \centering
    \begin{subfigure}{\columnwidth}
        \centering
        \includegraphics[width=\columnwidth]{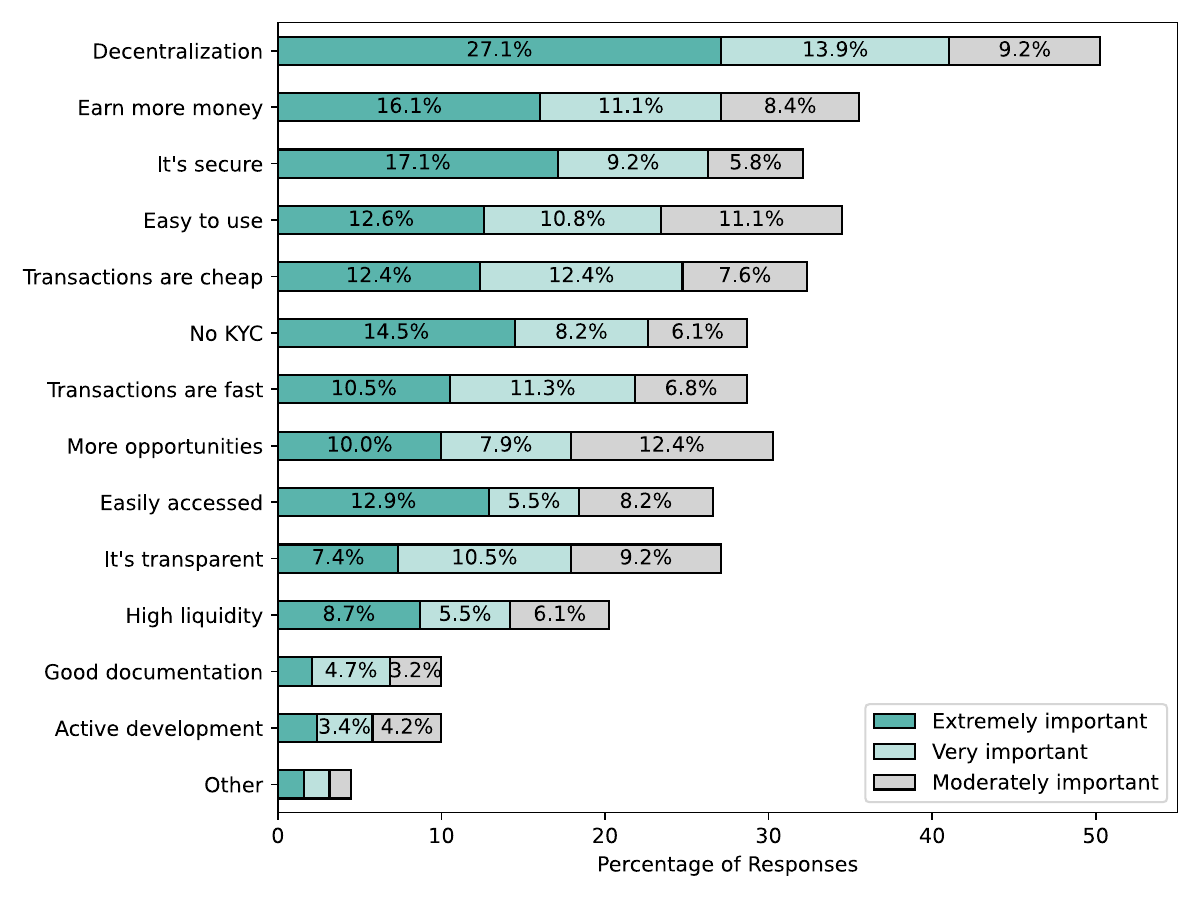}
        \caption{The survey results about love points of DeFi.}
        \label{fig:love-all}
    \end{subfigure}

    \smallskip

    \begin{subfigure}{\columnwidth}
        \centering
        \includegraphics[width=\columnwidth]{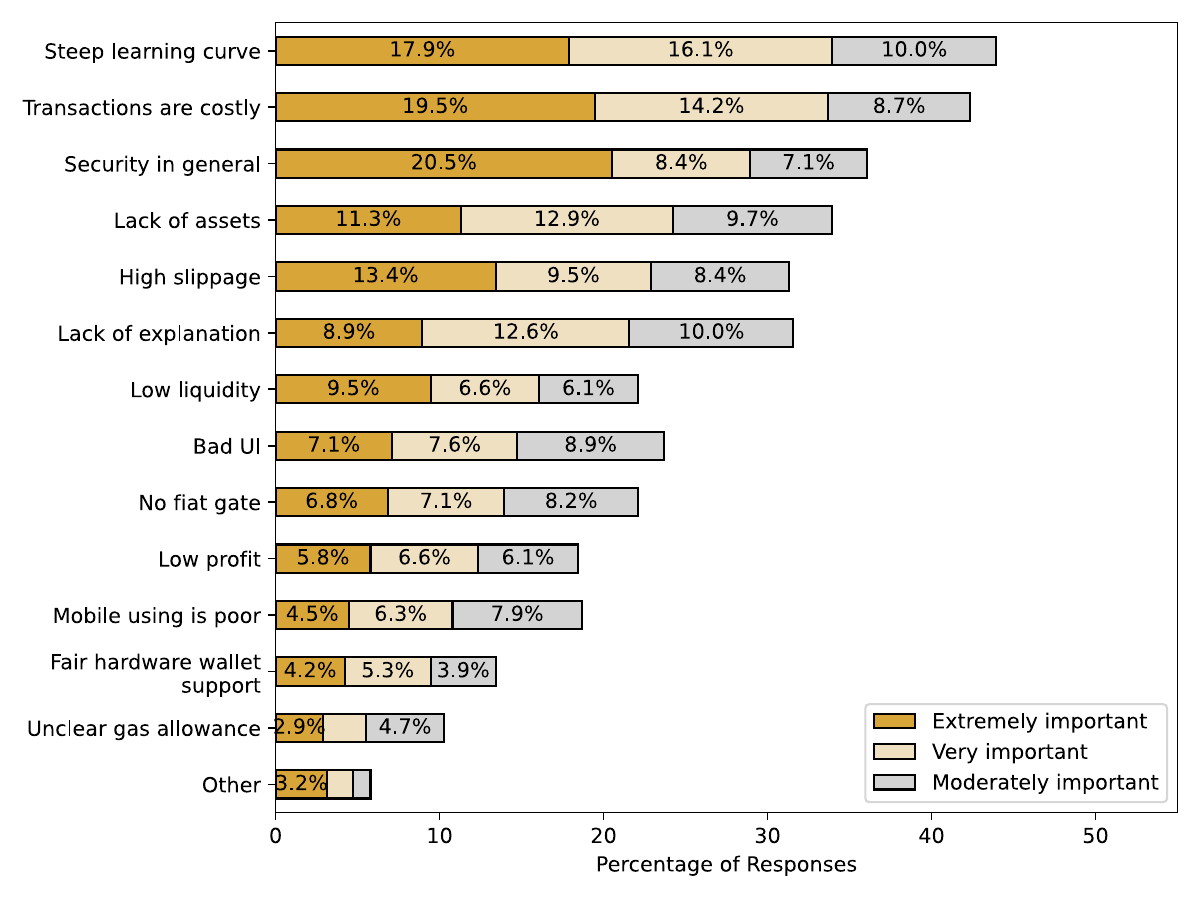}
        \caption{The survey results about pain points of DeFi.}
        \label{fig:pain-all}
    \end{subfigure}

    \caption{The survey results about love/pain points of DeFi.}
    \label{fig:love-pain}
\end{figure}

\subsubsection{Perceptions of DeFi risks and mitigation}
\label{s:qual-rq2}
This section explores \textbf{RQ2}, and reports details about DeFi users' security concerns and mitigation strategies.

\PP{Perceived risks}
With respect to the perceived risk questions, many participants expressed concerns about \texttt{smart contract exploitation}, some mentioned specific vulnerabilities: ``\textit{There are a lot of risks in the lending protocol like oracle price manipulation}'' (P7).
Several participants were concerned about cryptocurrency price volatility and associated \texttt{financial risks}. The third most frequently mentioned risk was \texttt{rug-pull}, a common pump-and-dump exit scam orchestrated by the project owners. P6 expressed concerns about the lack of regulations to protect DeFi users from rug-pulls: ``\textit{Without it being regulated by someone, there's always the risk of investing in a rug-pull project.}''

Other frequently noted risks include \texttt{theft of private key} from wallets, \texttt{instability of stablecoins} (risk of de-pegging from 1 USD), and \texttt{airdrop scam} related to malicious tokens stored in wallets, each of which was mentioned by a few participants. \texttt{Phishing} and \texttt{regulatory uncertainty} were each mentioned once.

\PP{Risk mitigation}
Participants reported mitigation strategies and security practices they have employed to mitigate the risks they were concerned about.
To deal with the risk of rug-pull and smart contract exploitation, most participants explained they perform thorough research on a given crypto project before investing. For instance, P9 said ``\textit{We do our research about the token [first]... and make sure it's a safe and reliable token before we buy [them].}''
To address the risks associated with private key compromise, some participants said they rely on hardware wallets to safeguard private keys. P7 explained ``\textit{A hacker can't take my crypto private key because I use a hardware wallet.}'' 

P4 and P6 explained that they regularly revoke token approvals to mitigate rug-pull risks. 
Token approval allows dApps to access users' wallets, and transfer tokens on users' behalf. This reduces users' re-approval efforts and txs costs.
However, token approval is often configured to allow unlimited token transfers, and such configurations could be exploited by malicious dApps to transfer the entire balance.
To that end, P4 shared thoughts on the risk of failing to revoke token approvals in a timely manner: ``\textit{If you leave them approved, you could have a bad upgrade [with] a vulnerability that might be exploited. And if you've approved unlimited amounts, then they can just spend all [of] your tokens...}''

\PP{Overconfidence in two-factor authentication}
P2 and P10 shared their strategy of using two-factor authentication (2FA) to mitigate many of the DeFi risks, including rug-pull and smart contract exploitation. P10 mentioned ``\textit{Two-factor authentication has been one of the best solutions for keeping wallets safe.}'' Although 2FA is an integral security practice for protecting user accounts in CeFi platforms, it is not supported on non-custodial (decentralized) wallets~\cite{metamask}. We report this as a critical security misconception: DeFi users being overly confident in the use of 2FA to protect their wallets, and, as a result, paying less attention to other important security practices.

\subsubsection{Real-world experience with DeFi breaches}
\label{s:qual-rq3}
To address \textbf{RQ3}, we study real-world victims' experiences and how they responded to a scam or hack.

\PP{Type of breaches.}
DeFi scams were reported more often than hacks by the interviewed victims. Among those who experienced DeFi hacks, several victims mentioned being affected by smart contract exploits, while some mentioned suffering from private key compromise. For example, P4, who is a dApp developer, shared an incident that involved the exploitation of a re-entrancy vulnerability in their smart contract:
``\textit{We lost million[s of dollars], it was not a good time... It was a re-entrancy vulnerability, and the audit completely missed it.}''

With respect to DeFi scam experiences, many participants encountered rug-pull incidents: ``\textit{He [the founder] deployed two smart contracts. One was fine [but] another was a rug-pull}'' (P7).
Several participants fell victim to phishing scams: ``\textit{I was DM-ed [direct messaged] by somebody who said, you just joined this channel, click this link to verify your account... I logged into my MetaMask... [and] half an hour later... everything was gone}'' (P12).

\PP{Response actions.}
We asked interview participants about the immediate actions taken after experiencing the last scam or hack incident.
Their responses include adopting new security practices, asking the support or development team for assistance, or not taking any action. Some victims mentioned they increased the net investment size, and moved to other DeFi services to quickly recover their loss. 
Among those who adopted a new security practice, P12 responded to a phishing scam by ``\textit{... disconnecting my wallet from every site, and revoking every [token] approval that I had out there.}'' Several victims who contacted the development team received no response, or a premature response that was not particularly helpful. P3, for instance, said ``\textit{I sent an email... but I did not get any reply, [and] I moved on.}'' P5 shared a similar experience: ``\textit{[They said] they will get back to me, but I never heard from them again.}''
One victim, who experienced rug-pull, contacted an IP lawyer in an attempt to sue the development team but stopped after learning about the low probability of getting anything back. Some of the victims, who were affected by phishing and rug-pull breaches, did not take any action.

\PP{Perception changes.}
Our objective was to understand how victims' security perceptions may have been affected by the breaches. 
We asked ``\textit{has your belief or perception of DeFi changed after experiencing the DeFi hack or scam}?''
Surprisingly, most interview participants explained their security perceptions of DeFi did not change; some even mentioned that their confidence in DeFi platforms increased despite the breach, praising the previously realized profits, and often blaming themselves for being careless.
P3, for instance, lost about 4,700 USD in a recent rug-pull incident but said ``\textit{my belief in cryptocurrency has grown stronger after [experiencing] that [DeFi scam] because I made good money from it... An opportunity to make money is something I believe in.}''
P9 blamed themselves: ``\textit{Oh, my belief did not change! I just felt like I was the victim of my own circumstance... Not doing enough research before diving into it... it was my fault basically.}'' P11 explained that it was a risk that they were willing to take for financial gain.

\PP{Blame distribution.}
Lastly, we analyzed the distribution of stakeholders that the victims held accountable for the experienced breaches. A few interview participants simply blamed hackers and scammers. About half of the participants blamed developers: ``\textit{It is the responsibility of the developers to spot their loophole [first]... and make amendments for investors' security}'' (P3).
Most participants held themselves accountable, and explained that they should have done more research prior to using a DeFi service.

\subsubsection{DeFi regulation preferences}
\label{s:qual-rq4}

This section delves into \textbf{RQ4} and investigates DeFi users' perceptions of regulations. 
The U.S. Securities and Exchange Commission (SEC) issued a statement~\cite{sec-defi-risk} in 2021 outlining ``pseudonymity'' and ``lack of transparency'' issues associated with DeFi.
In 2022, the SEC amended ``Rule 3b-16'' of the Exchange Act~\cite{sec-defi-proposal} to include all DeFi platforms in the ``exchange'' category. In the latest 2023 statement~\cite{sec-defi-2023}, the SEC explained that DeFi platforms need to conform to laws governing securities.

Given this background, our interview participants shared mixed feelings about regulation. Many participants shared positive feedback whereas the majority provided opposing feedback. Participants who endorsed DeFi regulation believed that it would promote DeFi security, reduce financial loss, and protect them from adversaries. 
P6, for example, explained ``\textit{A malicious user should be punished... there needs to be some justice...}''
Opposing participants were worried that regulation efforts sit uneasily with the decentralized and unregulated fundamentals of DeFi. Some were also concerned about regulations discouraging innovation, and tax implications.

%% file: quan-results.tex
\subsection{Online Survey Results}
\label{s:quan-results}

\subsubsection{Perceptions of DeFi}
\label{s:quan-rq1}
We focus on validating the interview results for \textbf{RQ1} in this section.
The survey was designed to distinguish exchanges and lending platforms as two different DeFi services.
However, because the user perceptions were not too different between the two services, we decided to simplify the analysis and report the aggregated results.

First, we derived and coded the aspects of users' enjoyment and frustration in DeFi, based on our interviews and a preliminary industry study~\cite{dexblue}.
From the initial pilot study, we noticed that people take too long and find it difficult to rank all given options based on relative importance. Hence, in the final survey design, we asked survey respondents to select and rank no more than three \textit{love/pain points}.
~\autoref{fig:love-pain} shows the results sorted by weighted scores, where ``Extremely important,'' ``Very important,'' and ``Moderately important'' represent 3, 2, and 1 scores, respectively.
The detailed results of the statistical tests
are presented in~\autoref{a:perception-test}.
Due to page limits, we only report and discuss highly ranked love/pain points.

\PP{Love points}
\texttt{Decentralization} was considered to be the most important love point. 
Our subsequent security analysis,
however, revealed that many participants have an inadequate understanding of decentralization and thus are optimistic about this characteristic contributing to security.
\texttt{Earn more money} ranked second, indicating that financial gains are important motivations for using DeFi services~\cite{dexblue}.

\PP{Pain points}
Participants regarded steep learning curves and high transaction (tx) costs, which are classified as \textit{usability}, and \textit{security} issues, as the most important pain points.
It's understandable that users deem DeFi tx fees expensive because, in addition to the inevitable gas fee\footnote{Blockchain miners' reward for executing transactions.}, dApps may also impose liquidity taker fee on tx to reward market creators who provide liquidity, and this fee varies among dApps.
In this case, some developers lowered their tx fees to offer a more cost-friendly environment~\cite{uniswap-v3,1inch-v5}.

\begin{figure}[t!]
    \centering
    \begin{subfigure}{\columnwidth}
        \centering
        \includegraphics[width=\columnwidth]{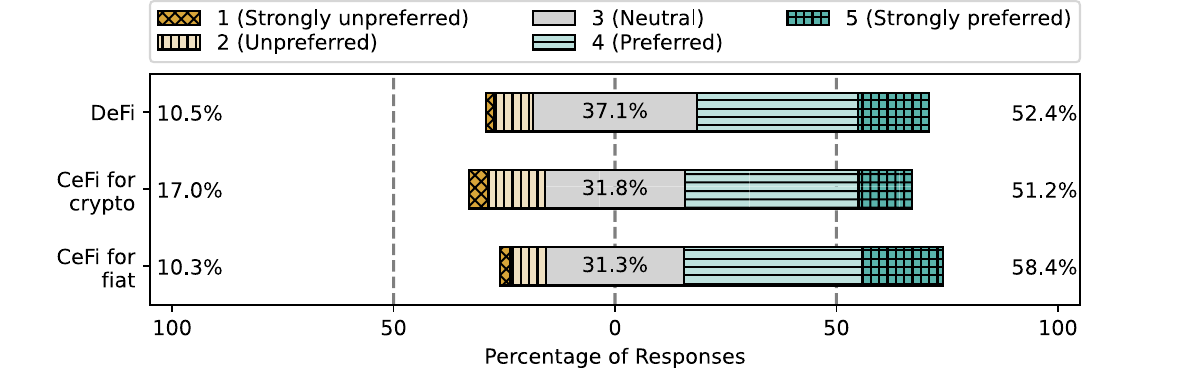}
        \caption{The comparison in terms of \textit{preference}}
        \label{fig:pref-all}
    \end{subfigure}

    \smallskip

    \begin{subfigure}{\columnwidth}
        \centering
        \includegraphics[width=\columnwidth]{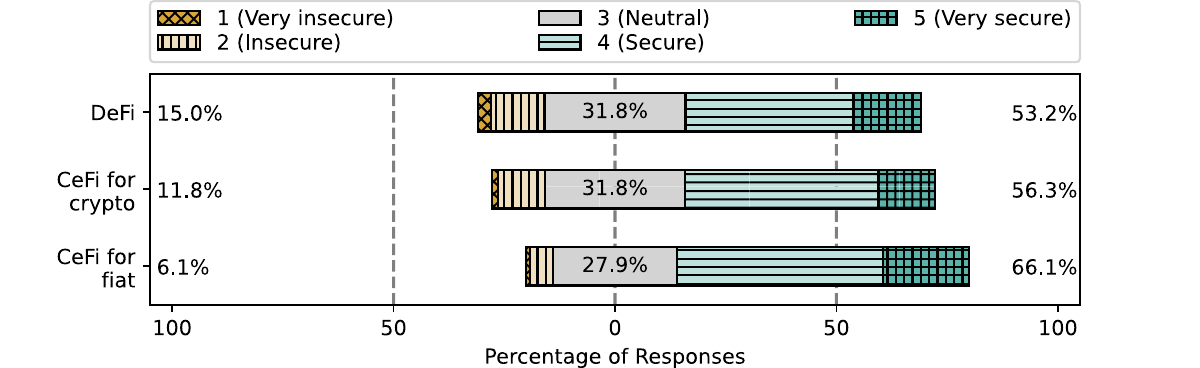}
        \caption{The comparison in terms of \textit{security}}
        \label{fig:sec-all}
    \end{subfigure}

    \smallskip

    \begin{subfigure}{\columnwidth}
        \centering
        \includegraphics[width=\columnwidth]{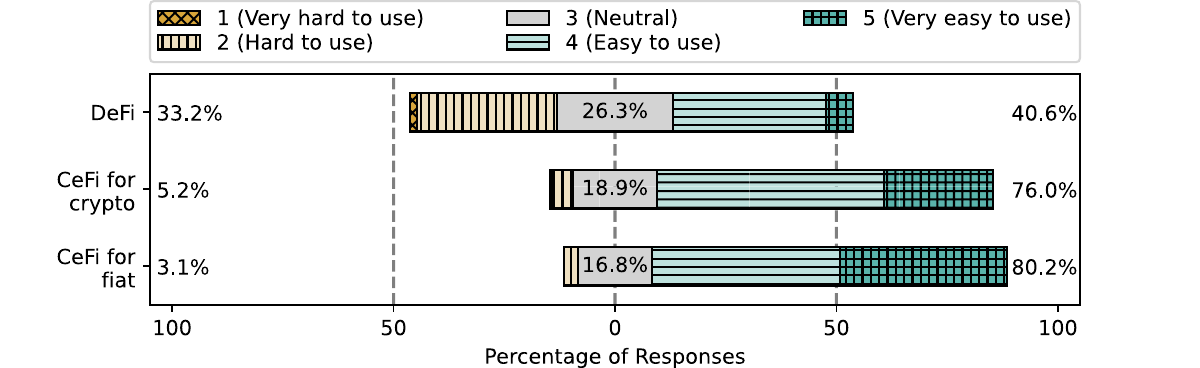}
        \caption{The comparison in terms of \textit{usability}}
        \label{fig:use-all}
    \end{subfigure}

    \smallskip

    \begin{subfigure}{\columnwidth}
        \centering
        \includegraphics[width=\columnwidth]{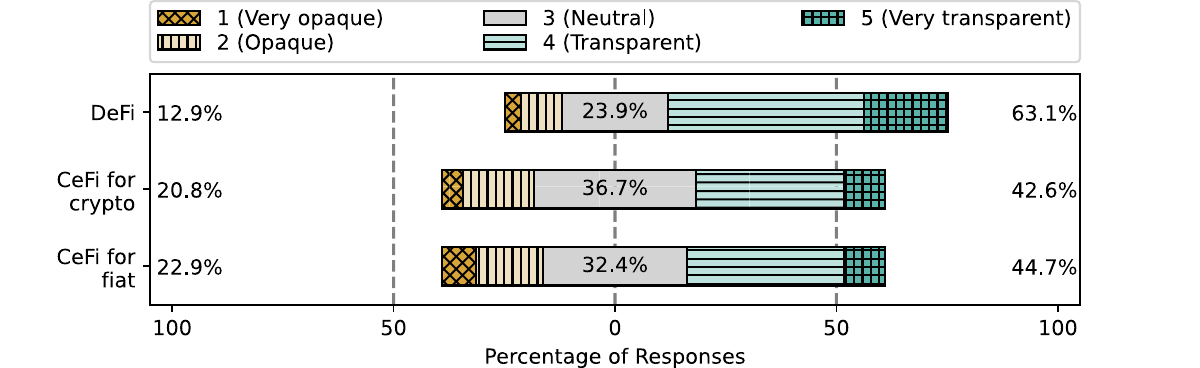}
        \caption{The comparison in terms of \textit{transparency}}
        \label{fig:trans-all}
    \end{subfigure}

    \smallskip

    \begin{subfigure}{\columnwidth}
        \centering
        \includegraphics[width=\columnwidth]{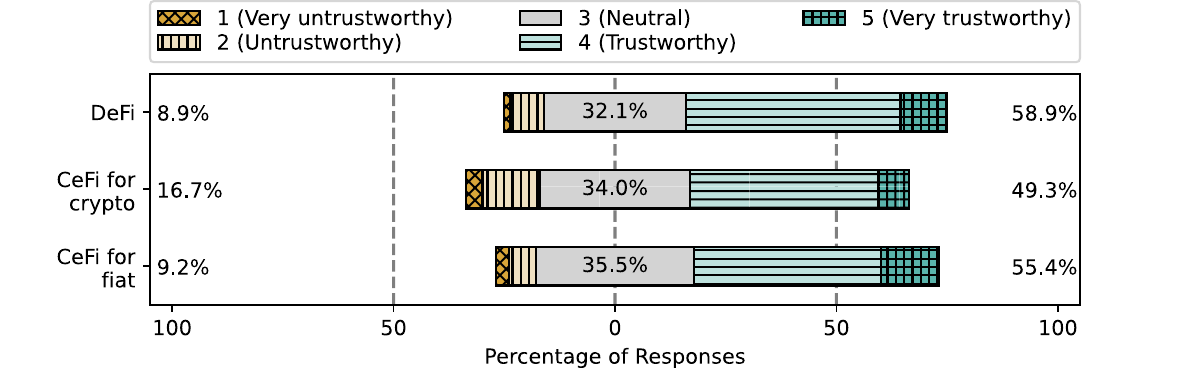}
        \caption{The comparison in terms of \textit{trust}}
        \label{fig:trus-all}
    \end{subfigure}

    \caption{The comparison of survey participants' perceptions between DeFi, CeFi for crypto, and CeFi for fiat.}
\end{figure}

\begin{table*}[!t]
\begin{threeparttable}
\centering
\begin{singlespacing}
    \tabcolsep=6pt
    \footnotesize
    \caption{Results showing significant differences in Mann-Whitney U tests under Bonferroni correction for DeFi's and CeFi's rate distributions of \textit{preference}, \textit{security}, \textit{usability}, \textit{transparency}, and \textit{trust}. The $p$-values and effect sizes in rank-biserial correlation $r$ are reported.}
    \label{t:perception-test}
\begin{tabular}{|c|c|c|c|c|c|c|c|c|c|c|}
\hline
\multirow{2}{*}{\textbf{Item}}& \multicolumn{2}{c|}{\textbf{\textit{Preference}}} & \multicolumn{2}{c|}{\textbf{\textit{Security}}} & \multicolumn{2}{c|}{\textbf{\textit{Usability}}} & \multicolumn{2}{c|}{\textbf{\textit{Transparency}}} & \multicolumn{2}{c|}{\textbf{\textit{Trust}}} \\
\cline{2-11}
& DeFi & CeFi\_C & DeFi & CeFi\_C & DeFi & CeFi\_C & DeFi & CeFi\_C & DeFi & CeFi\_C \\
\hline
\multirow{2}*{\textbf{CeFi\_C}\tnote{1}} & & \multirow{2}*{--}  & & \multirow{2}*{--} & $p$ = 2.2e-30 & \multirow{2}*{--} & $p$ = 1.1e-8 & \multirow{2}*{--} & $p$ = 8.4e-4 & \multirow{2}*{--} \\
 & & & & & $r$ = 0.46 & & $r$ = -- 0.23 & & $r$ = -- 0.13 & \\
\hline
\multirow{2}*{\textbf{CeFi\_F}\tnote{2}} & & $p$ = 9.0e-3 & $p$ = 4.4e-4 & $p$ = 2.1e-3 & $p$ = 6.0e-35 &  $p$ = 2.1e-3  &  $p$ = 3.1e-7 &   &   & $p$ = 1.1e-2 \\
 & & $r$ = 0.12 & $r$ = 0.15 & $r$ = 0.13 & $r$ = 0.55 & $r$ = 0.13 & $r$ = -- 0.23 & & & $r$ = 0.11 \\
\hline
\end{tabular}
\begin{tablenotes}
   \item $^1$ Abbreviated for \textit{CeFi for cryptocurrency}; $^2$ Abbreviated for \textit{CeFi for fiat currency}.
\end{tablenotes}
\end{singlespacing}
\end{threeparttable}
\end{table*}

Second, from the interview responses, we identified \textit{security}, \textit{usability}, \textit{transparency}, and \textit{trust} as the most common reasons (thematic codes) influencing users' \textit{preference} for DeFi and CeFi systems.
We report our Likert scale question results from the survey, which has been designed to investigate users' comparative perceptions.

\PP{Preference.}
In contrast to the interview responses, our survey results did not show a dominant preference for DeFi. Participants' preference rates are summarized in~\autoref{fig:pref-all}. 
We observed a statistically significant difference in preference rates between crypto CeFi and fiat CeFi ($p$ < 0.01) but the effect size was small ($|r|$ = 0.12).

\PP{Security.}
Unsurprisingly, ~\autoref{fig:sec-all} shows that users perceive DeFi as slightly less secure than CeFi. Statistical tests in~\autoref{t:perception-test} revealed significant differences in security perception between DeFi and fiat CeFi ($p$ < 0.001), and between crypto CeFi and fiat CeFi ($p$ < 0.01). The effect sizes were small in both cases ($|r|$ = 0.15 and $|r|$ = 0.13).

We also asked an open-ended question about the reasons for their perceived security ratings: 39.8\% considered DeFi to be secure due to its \texttt{decentralization} characteristic; this observation is well aligned with the love points explained before, 
in which the \texttt{it's secure} love point also recorded a high ranking.
However, some explanations were not particularly convincing. For example, one participant believed that ``\textit{decentralized [service] is always more safe than centralized [service]}'' (S512). Some participants mistakenly equated the decentralization of DeFi with that of the underlying blockchain and thus reported a misbelief about the security guarantees: ``\textit{[DeFi is] more secure because a hacker would have to override an entire blockchain [to steal funds in DeFi]}'' (S12).
\texttt{Self-custody of private keys} was also mentioned frequently (23.9\%). One participant explained ``\textit{[DeFi is] secure if the private keys are well stored}'' (S20), which is not the case. 
Some participants (19.3\%) considered DeFi to be secure simply due to \texttt{no hack or scam experience}.

Among those who considered DeFi to be insecure, 55.8\% of such participants mentioned \texttt{rampant hacks or scams} as the main reason. One participant explained ``\textit{I have [used] malicious smart contracts that steal your funds before}'' (S398).
\texttt{Decentralization} was the second most frequently mentioned reason (23.3\%) for feeling insecure.
For example, S224 explained ``\textit{[DeFi services] are easier to exploit than centralized [services].}''
Furthermore, some participants associated decentralization with a lack of regulations and controls, and emphasized such liberal aspects as a reason for feeling less confident about DeFi.

Taken together, we report that a significant portion of participants over-estimate DeFi's security due to their limited understanding of the concept of \texttt{decentralization}. Such participants do not seem to be aware of the new attack vectors introduced through the use of smart contracts~\cite{zhou:oakland:2023}.

\PP{Usability.}
The overall sentiment, as shown in~\autoref{fig:use-all}, was that both crypto and fiat CeFi are easier to use than DeFi ($p$ < 0.001 in both cases). Near-large and large effect sizes further emphasize the magnitude of differences ($|r|$ = 0.46 and $|r|$ = 0.55) which was somewhat unexpected since our interview results revealed some participants prefer DeFi due to its fast transaction and interoperability benefits (See \autoref{s:qual-rq1}).

We asked survey participants to explain the usability scores they assigned to DeFi. Among 60.1\% of explanations that provided negative feedback,
40.2\% mentioned \texttt{steep learning curve}. Many participants felt overwhelmed by the new technologies they had to learn and use, including the concept of dApps, blockchains, and the interactions with non-custodial wallets. 
11.8\% of such participants explained \texttt{it's complex}, and 8.3\% mentioned the learning curve associated with \texttt{blockchain knowledge required}. However, among those negative feedback, a noticeable 14.4\% mentioned that using DeFi was \texttt{easy after education}.
These observations indicate that users may struggle initially and face various learning challenges but through continued use and some educational support, DeFi could become an easier platform to use.

\PP{Transparency.}
We defined \textit{transparency} in our survey as the extent to which a subject discloses operational and transactional details.
Evidence gathered through ~\autoref{fig:trans-all} and~\autoref{t:perception-test} validates that participants perceive DeFi to be more transparent: distribution of perceived transparency differed significantly between DeFi and other two CeFi services ($p$ < 0.001 in both cases), demonstrating small to medium effect sizes ($|r|$ = 0.23 in both cases).

\PP{Trust.}
Survey results, summarized in~\autoref{fig:trus-all} and~\autoref{t:perception-test}, indicate that users consider crypto CeFi to be the least trustworthy platform -- its perceived trust level distribution showed statistically significant differences compared to both DeFi ($p$ < 0.001, $|r|$ = 0.13) and fiat CeFi ($p$ < 0.05, $|r|$ = 0.11).

\subsubsection{Perceptions of DeFi risks and mitigation}
\label{s:quan-rq2}
This section presents large-scale validation for \textbf{RQ2}, focusing on DeFi users' security concerns and preventive measures.

\begin{figure}[t!]
    \centering
    \includegraphics[width=\columnwidth]{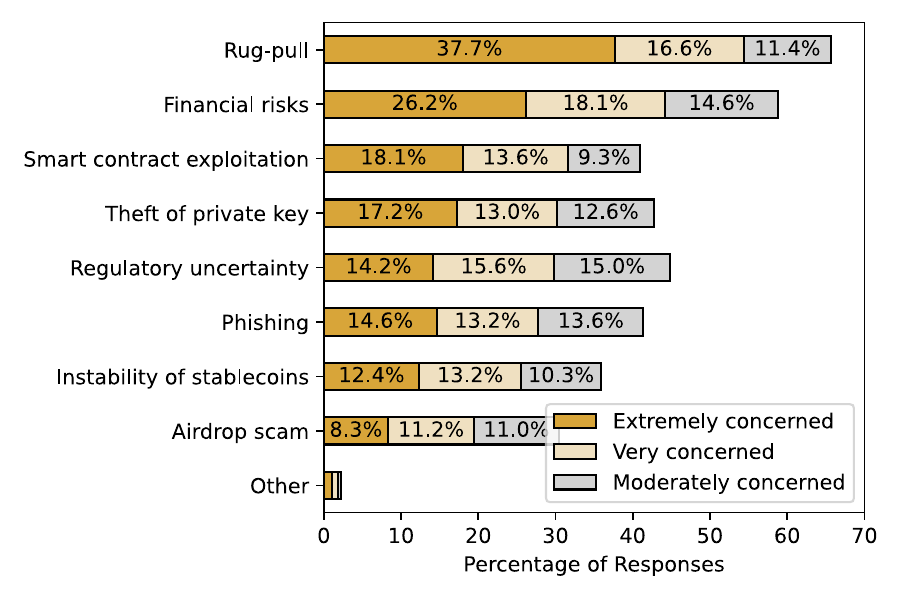}
    \caption{The DeFi risks that concern survey participants.}
    \label{fig:risk-all}
\end{figure}

\begin{figure}[t!]
\centering
    \adjustbox{trim=0 15 10 40, clip}{
        \includegraphics[width=\columnwidth]{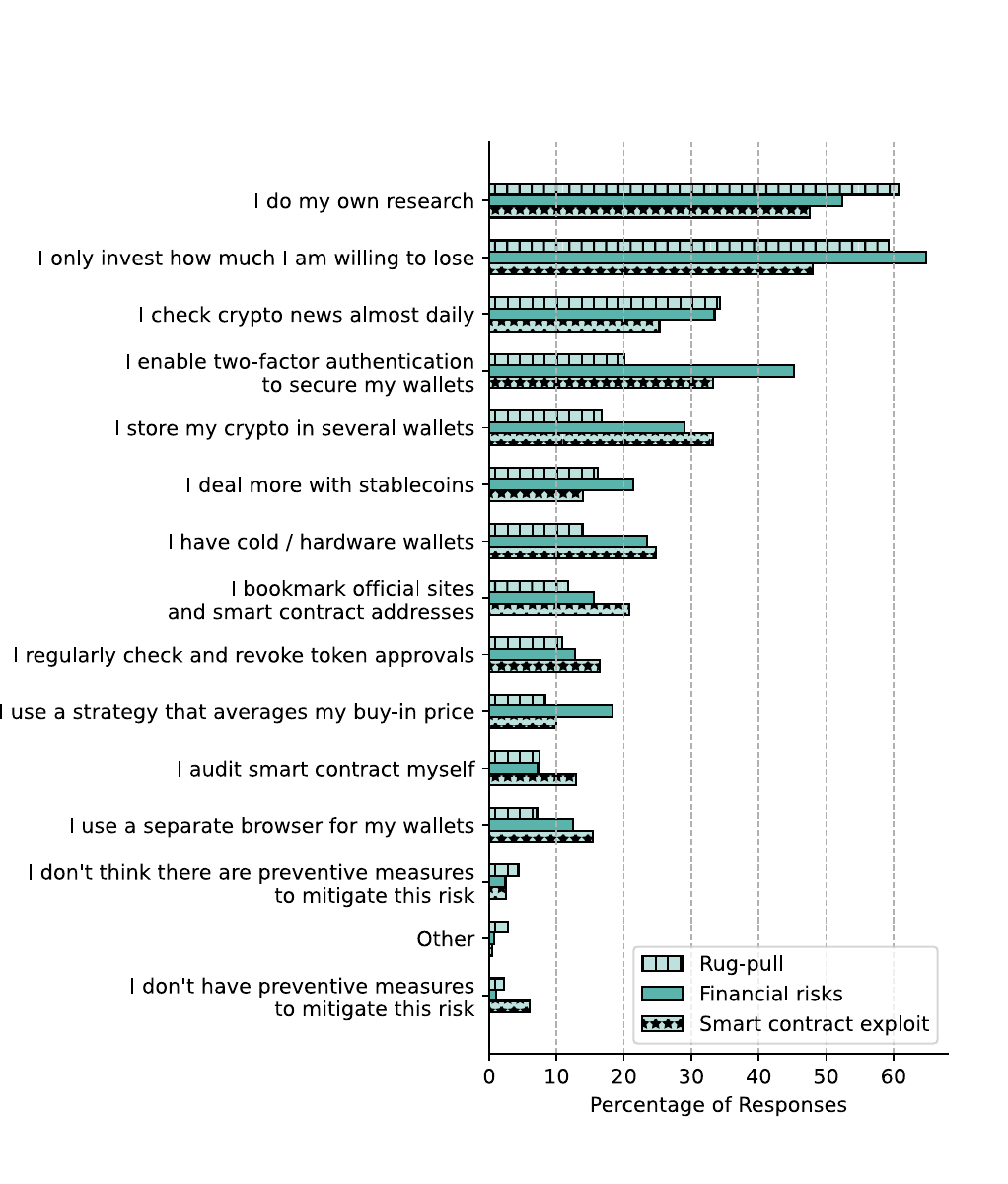}
    }
    \caption{The survey results about how they mitigate \textit{rug-pull}, \textit{financial risks}, and \textit{smart contract exploitation}.}
    \label{fig:risk-mitigate-top-3}
\end{figure}

\begin{table*}[t!]
    \centering

    \tabcolsep=3pt
    \footnotesize
    \caption{The preventive measures adopted by survey participants against \textit{rug-pull}, \textit{financial risk}, and \textit{smart contract exploitation}.}
    \label{t:risk-mitigate-top-3-victim}
    \begin{tabular}{l|crcr|crcr|crcr}
    \hlineB{3}
    \multirow{3}{*}{\textbf{Preventive Measures}} & \multicolumn{4}{c|}{\textbf{Rug-pull}} & \multicolumn{4}{c|}{\textbf{Financial Risk}} & \multicolumn{4}{c}{\textbf{Smart Contract Exploit}} \\
    \cline{2-13}
                                         & \multicolumn{2}{c}{Non-victim} & \multicolumn{2}{c|}{Victim} & \multicolumn{2}{c}{Non-victim} & \multicolumn{2}{c|}{Victim} & \multicolumn{2}{c}{Non-victim} & \multicolumn{2}{c}{Victim} \\
                                         & \multicolumn{2}{c}{$N$=250} & \multicolumn{2}{c|}{$N$=74} & \multicolumn{2}{c}{$N$=237} & \multicolumn{2}{c|}{$N$=53} & \multicolumn{2}{c}{$N$=138} & \multicolumn{2}{c}{$N$=64} \\
    \hlineB{3}
    \texttt{I do my own research}   & 155  & (62.0\%)  & 42  & (56.8\%)  & 118   & (49.8\%)  & 34  & (64.2\%)  & 59  & (42.8\%)  & 37 & (57.8\%)  \\
    \texttt{I only invest how much I am willing to lose}   & 144  & (57.6\%)  & 48  & (64.9\%)  & 156   & (65.8\%)  & 32  & (60.4\%)  & 59  & (42.8\%)  & 38 & (59.4\%)  \\
    \texttt{I check crypto news almost daily}   & 85  & (34.0\%)  & 26  & (35.1\%)  & 83   & (35.0\%)  & 14  & (26.4\%)  & 38  & (27.5\%)  & 13 & (20.3\%)  \\
    \texttt{I enable two-factor authentication to} & \multirow{2}{*}{49} & \multirow{2}{*}{(19.6\%)} & \multirow{2}{*}{16} & \multirow{2}{*}{(21.6\%)} & \multirow{2}{*}{106} & \multirow{2}{*}{(44.7\%)} & \multirow{2}{*}{25} & \multirow{2}{*}{(47.2\%)}  & \multirow{2}{*}{45}  & \multirow{2}{*}{(32.6\%)}  & \multirow{2}{*}{22} & \multirow{2}{*}{(34.4\%)}  \\
    \hspace{3mm}\texttt{secure my wallets} &  &   &   &  &   &  &  &  &  &   & &   \\
    \texttt{I deal more with stablecoins}   & 44  & (17.6\%)  & 8  & (10.8\%)  & 53   & (22.4\%)  & 9  & (17.0\%)  & 16  & (11.6\%)  & 12 & (18.8\%)  \\
    \texttt{I store my crypto in several wallets}   & 39  & (15.6\%)  & 15  & (20.3\%)  & 69   & (29.1\%)  & 15  & (28.3\%)  & 39  & (28.3\%)  & 28 & (43.8\%)  \\
    \texttt{I have cold/hardware wallets}   & 31  & (12.4\%)  & 14  & (18.9\%)  & 61   & (25.7\%)  & 7  & (13.2\%)  & 33  & (23.9\%)  & 17 & (26.6\%)  \\
    \texttt{I bookmark official sites and}   & \multirow{2}{*}{26}  & \multirow{2}{*}{(10.4\%)}  & \multirow{2}{*}{12}  & \multirow{2}{*}{(16.2\%)}  & \multirow{2}{*}{37}   & \multirow{2}{*}{(15.6\%)}  & \multirow{2}{*}{8}  & \multirow{2}{*}{(15.1\%)}  & \multirow{2}{*}{28}  & \multirow{2}{*}{(20.3\%)}  & \multirow{2}{*}{14} & \multirow{2}{*}{(21.9\%)}  \\
    \hspace{3mm}\texttt{smart contract addresses}  &  &   &   &  &   &  &  &  &  &   & &   \\
    \texttt{I regularly check and revoke token approvals}   & 22  & (8.8\%)  & 13  & (17.6\%)  & 27   & (11.4\%)  & 10  & (18.9\%)  & 20  & (14.5\%)  & 13 & (20.3\%)  \\
    \texttt{I use a strategy that averages my buy-in price}   & 18  & (7.2\%)  & 9  & (12.2\%)  & 37   & (15.6\%)  & 16  & (30.2\%)  & 15  & (10.9\%)  & 5 & (7.8\%)  \\
    \texttt{I audit smart contract myself}   & 16  & (6.4\%)  &  8 & (10.8\%)  & 15   & (6.3\%)  & 6  & (11.3\%)  & 19  & (13.8\%)  & 7 & (10.9\%)  \\
    \texttt{I use a separate browser for my wallets}   & 16  & (6.4\%)  & 7  & (9.5\%)  & 28   & (11.8\%)  & 8  & (15.1\%)  & 20  & (14.5\%)  & 11 & (17.2\%)  \\
    \texttt{I don't think there are preventive measures to}   & \multirow{2}{*}{10}  & \multirow{2}{*}{(4.0\%)}  & \multirow{2}{*}{4}  & \multirow{2}{*}{(5.4\%)}  & \multirow{2}{*}{7}   & \multirow{2}{*}{(3.0\%)}  & \multirow{2}{*}{0}  & \multirow{2}{*}{(0\%)}  & \multirow{2}{*}{5}  & \multirow{2}{*}{(3.6\%)}  & \multirow{2}{*}{0} & \multirow{2}{*}{(0\%)}  \\
    \hspace{3mm}\texttt{mitigate this risk}   &  &   &   &  &   &  &  &  &  &   & &   \\
    \texttt{I don't have preventive measures to}   & \multirow{2}{*}{6}  & \multirow{2}{*}{(2.4\%)}  & \multirow{2}{*}{1}  & \multirow{2}{*}{(1.4\%)}  & \multirow{2}{*}{3}   & \multirow{2}{*}{(1.3\%)}  & \multirow{2}{*}{0}  & \multirow{2}{*}{(0\%)}  & \multirow{2}{*}{10}  & \multirow{2}{*}{(7.2\%)}  & \multirow{2}{*}{2} & \multirow{2}{*}{(3.1\%)}  \\
    \hspace{3mm}\texttt{mitigate this risk} &  &   &   &  &   &  &  &  &  &   & &   \\
    \texttt{Other}   & 5  & (2.0\%)  & 4  & (5.4\%)  & 2   & (0.8\%)  & 0  & (0\%)  & 0  & (0\%)  & 1 & (1.6\%)  \\
    \hlineB{3}
    \end{tabular}
\end{table*}

\PP{Perceived risks.}
The survey questions about risks were constructed based on those frequently mentioned codes. We asked survey participants to select three most concerning risks and rank them based on concern levels. 
The concern level distributions (ordered by weighted scores) are shown in \autoref{fig:risk-all}, and the statistical significance between the distributions is measured and reported in \autoref{a:risk-test}.
In line with the interview findings, \texttt{rug-pull}, \texttt{financial risks}, and \texttt{smart contract exploitation} were the top three DeFi risks that survey participants were concerned about.

\PP{Risk mitigation.}
\autoref{fig:risk-mitigate-top-3} presents the security practices commonly employed by our survey participants to mitigate the originally reported three security concerns. First, in line with the interview responses, the most commonly practiced strategy involved \texttt{I do my own research} and \texttt{I only invest how much I am willing to lose}, which are somewhat general investment strategies. Second, token approvals are a critical attack vector. Yet, the adoption rate of the most appropriate countermeasure, \texttt{I regularly check and revoke token approvals}, was only practiced by 10.8\% of participants who shared their concerns about rug-pull scams. Although 2FA is not suitable for preventing rug-pulls, financial loss, and smart contract exploits, participants reported high adoption rates of 2FA for those three risks: recording 20.1\%, 45.2\%, and 33.2\% adoption rates. Among participants who adopted at least one \textit{technical} solution (2FA, hardware wallet, or revoking token approvals) for each risk, 62.4\%, 80.4\%, and 65\% were using 2FA, respectively. Among such 2FA users, 57.1\%, 56.5\%, and 49.3\% were using 2FA as the \emph{only} technical countermeasure -- indicating that the majority may be overconfident in the security guarantees offered by 2FA.

\PP{Influence of DeFi incidents}
We also investigated whether falling victim to DeFi hacks or scams affects users' security perceptions and practices. 
\autoref{t:risk-mitigate-top-3-victim} compares practices employed by non-victims and victims to mitigate their top three concerns. Noticeably, the victims' adoption rate for ``revoking token approvals'' was slightly higher in all three risks. Victims used the \texttt{I store my crypto in several wallets} strategy more frequently than non-victims to prevent rug-pull and smart contract exploits.
Hence, there may be some tendency for victims to become more vigilant. Chi-squared tests, however, did not show significant differences between the two groups in all three risks ($p=0.51$, $p=0.23$, and $p=0.46$).

There were also a few concerning trends: integral practices such as timely checking and revoking token approvals and opting for hardware wallets were not sufficiently adopted by victims. Further, the misconception about 2FA being adequate in mitigating the top three concerns was also prevalent among the victims.

\subsubsection{Real-world experience with DeFi breaches}
\label{s:quan-rq3}

\begin{table}[t!]
    \centering
    \tabcolsep=1.5pt
    \footnotesize
    \caption{DeFi Hack type encountered by survey participants, and whether they had a loss to that incident.}
    \label{t:hack-type}
    \begin{tabular}{lrrrr}
    \hlineB{3}
    \multicolumn{1}{c}{\textbf{DeFi Hack Type}} & \multicolumn{2}{c}{\textbf{Occurrence}} & \multicolumn{2}{c}{\textbf{Loss}}\\
    \hlineB{3}
    \texttt{Smart contract exploitation} & 22 & (37.9\%)  & 14 & (63.6\%)\\
    \hline
    \texttt{Cross-chain bridge attack} & 11 & (19.0\%)  & 9 & (81.8\%) \\
    \hline
    \texttt{Theft of private key for} & \multirow{2}{*}{9} & \multirow{2}{*}{(15.5\%)} & \multirow{2}{*}{6} & \multirow{2}{*}{(66.7\%)} \\
    \hspace{3mm}\texttt{my own wallets} &  &  &  &  \\
    \hline
    \texttt{Protocol front-end attack} & 6 & (10.3\%) & 5 & (83.3\%) \\
    \hline
    \texttt{Theft of private key for} & \multirow{2}{*}{4} & \multirow{2}{*}{(6.9\%)} & \multirow{2}{*}{2} & \multirow{2}{*}{(50.0\%)}\\
    \hspace{3mm}\texttt{protocol smart contract} &  &  &  & \\
    \hline
    \texttt{Other} & 3 & (5.2\%) & 1 & (33.3\%)\\
    \hline
    \texttt{I have no idea} & 3 & (5.2\%) & 1 & (33.3\%)\\
    \hlineB{3}
    \end{tabular}
\end{table}

\begin{table}[t!]
    \centering
    \tabcolsep=2pt
    \footnotesize
    \caption{DeFi Scam type encountered by survey participants, and whether they had a loss to that incident.}
    \label{t:scam-type}
    \begin{tabular}{lrrrr}
    \hlineB{3}
    \multicolumn{1}{c}{\textbf{DeFi Scam Type}} & \multicolumn{2}{c}{\textbf{Occurrence}} & \multicolumn{2}{c}{\textbf{Loss}}\\
    \hlineB{3}
    \texttt{Rug-pull} & 40 & (54.8\%)  & 31 & (77.5\%)\\
    \hline
    \texttt{Phishing} & 16 & (21.9\%)  & 8 & (50.0\%) \\
    \hline
    \texttt{Wallet dusting/Airdrop scam} & 9 & (12.3\%) & 5 & (55.6\%) \\
    \hline
    \texttt{Stablecoin meltdown} & 4 & (5.5\%) & 2 & (50.0\%) \\
    \hline
    \texttt{Other} & 3 & (4.1\%) & 3 & (100\%) \\
    \hline
    \texttt{I have no idea} & 1 & (1.4\%) & 0 & (0\%)\\
    \hlineB{3}
    \end{tabular}
\end{table}

\begin{table*}[t!]
    \centering
    \tabcolsep=2pt
    \footnotesize
    \caption{The number of survey participants choosing each reason for unchanged or strengthened belief in DeFi after the incidents.}
    \label{t:belief-not-decrease-reason}
    \begin{tabular}{lrr|lrr}
    \hlineB{3}
    \multicolumn{3}{c|}{\textbf{DeFi Hack}} & \multicolumn{3}{c}{\textbf{DeFi Scam}} \\
    \hlineB{3}
    \multicolumn{1}{c}{Reason} & \multicolumn{2}{c|}{Count} & \multicolumn{1}{c}{Reason} & \multicolumn{2}{c}{Count} \\
    \hline
    \texttt{It was a risk that I was willing to take} & 16 & (55.2\%) & \texttt{It was a risk that I was willing to take} & 23 & (59.0\%) \\
    \hline
    \texttt{I am making profits regularly from DeFi} & 12 & (41.4\%) & \texttt{It was my fault, I could have been smarter} & 21 & (53.8\%) \\
    \hline
    \texttt{It was my fault, I could have been smarter} & 11 & (37.9\%) & \texttt{The technology outweighs the downside of scams} & 17 & (43.6\%) \\
    \hline
    \texttt{I made more research after the hack} & 11 & (37.9\%) & \texttt{I made more research after the scam} & 15 & (38.5\%) \\
    \hline
    \texttt{The technology outweighs the downside of hacks} & 10 & (34.5\%) & \texttt{I can use trustworthy DeFi apps} & 14 & (35.9\%) \\
    \hline
    \texttt{I can use trustworthy DeFi apps} & 9 & (31.0\%) & \texttt{I am making profits regularly from DeFi} & 13 & (33.3\%) \\
    \hline
    \texttt{I am a full-time DeFi trader} & 3 & (10.3\%) & \texttt{Other} & 2 & (5.1\%) \\
    \hline
    \texttt{Other} & 0 & (0\%) & \texttt{I am a full-time DeFi trader} & 0 & (0\%) \\
    \hlineB{3}
    \end{tabular}
\end{table*}

\begin{figure}[t!]
\centering
    \adjustbox{trim=2 2 20 25, clip}{
        \includegraphics[width=\columnwidth]{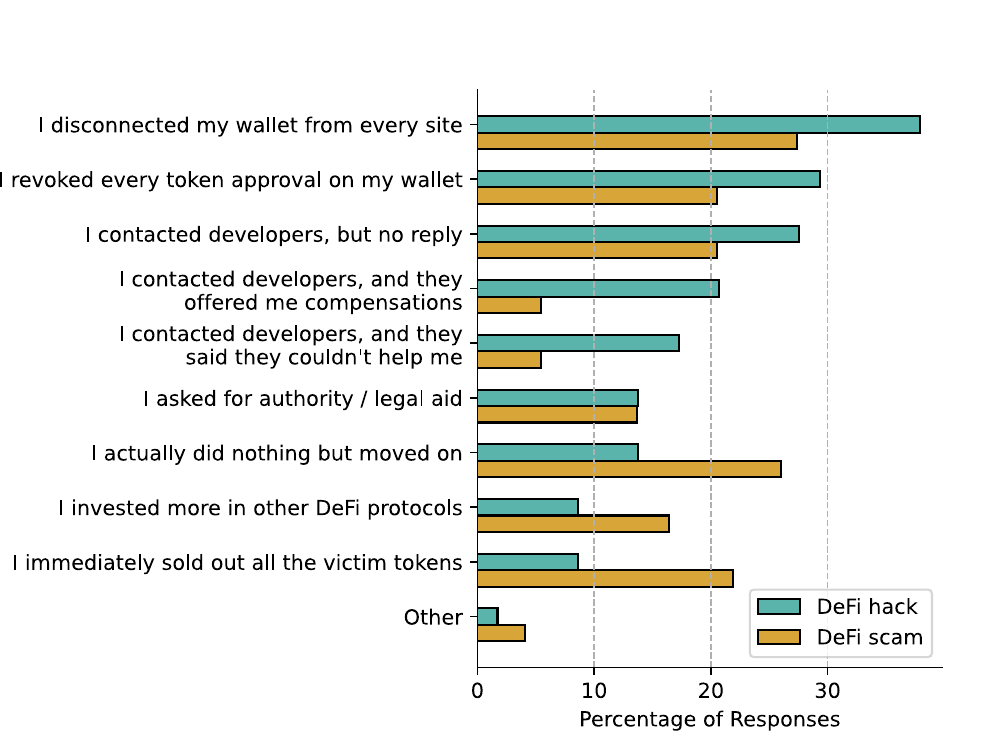}
    }
    \caption{The actions that victims in survey participants took after the latest DeFi incident that they experienced.}
    \label{fig:action}
\end{figure}

\begin{figure}[t!]
\centering
    \adjustbox{trim=10 5 0 15, clip}{
        \includegraphics[width=\columnwidth]{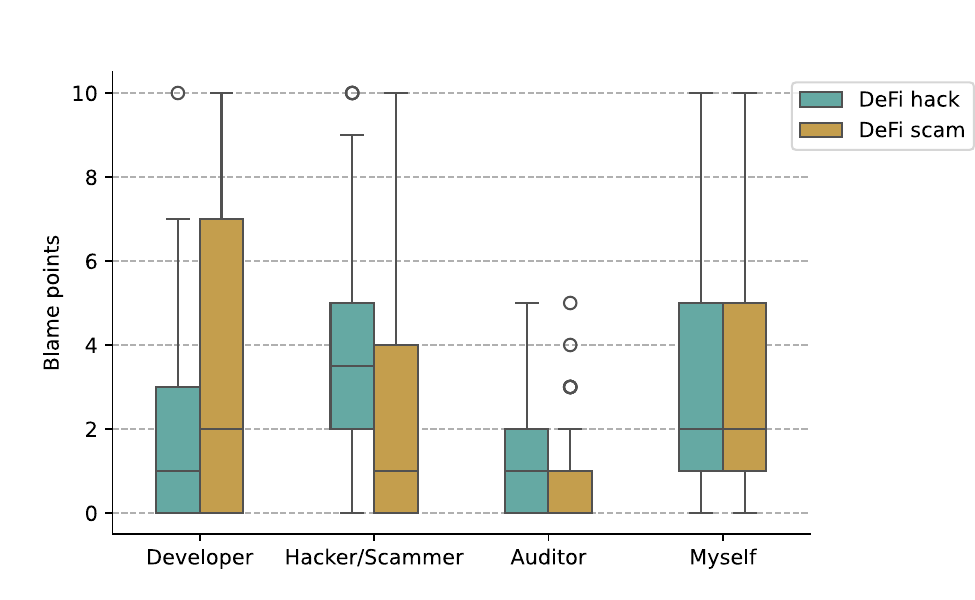}
    }
    \caption{
    The blame distribution by victims in the latest DeFi incident that they experienced.
    }
    \label{fig:blame-points}
\end{figure}

To address \textbf{RQ3}, we validated the interview results on real-world victims' experiences, perceptions, and responses to scams or hacks.

\PP{Being affected multiple times.}
In the online survey, we defined a DeFi hack as ``\textit{the act of identifying and then exploiting vulnerabilities in the DeFi domain}'', and provided unauthorized wallet access and dApp exploit as two examples.
11.8\% of the participants reported having previously experienced a DeFi hack at least once -- we refer to such participants as ``victims''. 
We then inquired victims about the number of times they have experienced DeFi hacks: the mean number of hacks was 1.8 ($\sigma$ = 1.6).
Similarly, we defined a DeFi scam as ``\textit{the act of tricking users or pretending to be a valid dApp, and stealing users' crypto assets}'', and hinted that common DeFi scams include transferring malicious cryptocurrency to victims' wallets
or being affected by rug-pulls.
In line with the interview results, 
more participants seem to have fallen victim to DeFi scams than hacks: 14.8\% of participants reported having experienced DeFi scams at least once. Surprisingly, the mean number of experienced scams was 4.7 ($\sigma$ = 12.5). Worryingly, some participants mentioned experiencing more than 10 scams in the past. 
We contacted the victims again to confirm those numbers.

\PP{Type of breaches.}
Victims reported the category of the DeFi hack they recently fell into, which is presented in \autoref{t:hack-type}.
\texttt{Smart contract exploitation} was the most prevalent attack, experienced by 22 victims.
9 victims suffered private key theft on their wallets. Other breaches are related to different vulnerabilities as discussed in \autoref{s:bkg}.
Regarding DeFi scams, \autoref{t:scam-type} confirms that \texttt{rug-pull} is the most dominant scam: more than half of the victims were affected. \texttt{Phishing} ranked second. Despite the prevalence of phishing attacks and the number of affected victims, the overall concern level for phishing (as reported in \autoref{fig:risk-all}) was much lower than other risks such as financial risk or private key theft. We surmise this may be due to the fact that phishing incidents do not always result in a direct financial loss: indeed, just 50.0\% of \texttt{phishing} victims reported losing money as a result of the incident; in comparison, a significantly larger proportion of victims reported financial implications in the case of \texttt{rug-pull} (77.5\%) and \texttt{smart contract exploitation} (63.6\%).

\PP{Response actions.}
\autoref{fig:action} presents a summary of the victims' response actions and the frequency of occurrence.
The most frequent response action was ``disconnect wallet'' and ``revoke token approval'' -- however, their overall adoption rates were just 27.4\% and 20.5\% after scams, respectively. Worryingly, 13.8\% of hack victims and 26.0\% of scam victims failed to take any action. These findings are in line with the observations reported in \autoref{s:quan-rq2}:
users generally lack knowledge about appropriate security practices that need to be considered after being affected by a breach. 
Another concerning trend entails several victims (16.4\% after scams) increasing the net investment amount, and using other DeFi services to quickly recover their loss. Such victims may not have put in any effort to improve their security practices even after experiencing various hacks and scams, including rug-pull (23.5\%)
and smart contract exploitation (11.8\%), 
and merely focused on finding new DeFi services to generate profit again.

\PP{Perception changes.}
The survey results confirmed interview observations at a larger scale: 50.0\% of hack victims and 53.4\% of scam victims indicated that their security perceptions did not change or confidence increased despite experiencing the hack or scam.
Their reasons are summarized in \autoref{t:belief-not-decrease-reason}.
For both scams and hacks, ``willingness to take risks'' was the most prevalent reason for unchanged perceptions, followed by ``making regular profits'' and ``self-blaming''. Taken together, these results suggest that DeFi users' financial motivations are very strong, and previous experiences and educational support may have minimal impact on improving their security practices.

\PP{Blame distribution.}
For validation, we asked our survey participants to compose blame points, with a sum of ten, for developers, hackers/scammers, auditors, and themselves, regarding the most recent incident they fell into. The survey results are summarized in \autoref{fig:blame-points}.
In line with the interview results, self-blame was prevalent in both incidents. Even the victims held themselves more accountable than developers for being affected by a DeFi hack.
We observed a very wide interquartile blame range for developers in the scam scenario: this may be due to the prevalence of \texttt{rug-pull} experiences, in which the developers often play the role of scammers.

\subsubsection{DeFi regulation preferences}
\label{s:quan-rq4}

The survey responses confirmed interview observations. 46.9\% of the participants disagreed with the statement ``\textit{I am in favor of DeFi being regulated by authorities.}'' 35.3\% agreed, and the rest were neutral. The top three reasons for endorsing regulation were \texttt{contribute to making DeFi secure} (46.2\%), \texttt{less financial loss for users} (37.7\%), and \texttt{penalize nefarious persons} (34.3\%). In comparison, the most prevalent reason for opposing regulation was \texttt{regulation brings in taxes} (53.1\%), followed by \texttt{regulation hinders innovation} (48.5\%) and \texttt{nefarious persons' misconduct still} (34.3\%).

We also examined how the number of adopted mitigation techniques (RQ2) and DeFi scam/hack experiences (RQ3) influence users' regulation preferences. The first investigation involved dividing participants into three groups for each of the top three risks: those who did not adopt any mitigation technique, those who adopted one or two techniques, and those who adopted more than two. We did not, however, find any statistically significant difference in the regulation preference rates between those three groups. The second investigation involved comparing regulation preference rates between victims and non-victims: again, we did not identify any significant difference between the two groups. 

Defining an adequate level of regulation is a complicated issue, and warrants further investigation. However, considering that security risks are often being ignored by DeFi users due to their strong financial motivations, and the most concerning implication of regulation is related to paying additional tax (this is a civil obligation anyway), we carefully hint toward developing an appropriate level of regulations to mandate strong security practices, and help victims protect their assets. 

%% file: takeaways.tex
\subsection{Key Takeaways}
\label{s:takeaways}

In this section, we highlight the key findings based on the evidence gathered from the two studies. First, profitability and decentralization, as one would expect, are the two key factors contributing to users' preference for DeFi. Users often blindly believe that decentralization alone translates to strong security and reliability. However, despite this common misconception, users still perceive DeFi as less secure compared to other CeFi services (See \autoref{s:qual-rq1} and \autoref{s:quan-rq1}). Second, similar to security behaviors observed in cryptocurrency systems~\cite{abramova:chi:2021}, DeFi users do not employ adequate security controls to mitigate their top concerns. Alarmingly, DeFi users appear overly confident that 2FA will effectively protect them from common DeFi threats. In contrast to the victim characteristics reported in~\cite{chen2017securing}, DeFi victims did not show significant improvement in employed security practices compared to non-victims, implying that experience and education may have a limited impact (See \autoref{s:qual-rq2} and \autoref{s:quan-rq2}). Third, \texttt{rug-pull} was the most prevalent breach experienced by the victims, and this observation aligns with the overall concern levels and the findings in~\cite{cernera:sec:2023}. \texttt{Phishing} was another prevalent attack among the victims, but the perceived concern levels were ranked among the bottom three risks. This contrasting trend may be explained by the fact that phishing does not always result in a direct financial loss (See \autoref{s:qual-rq3} and \autoref{s:quan-rq3}). Fourth, contrary to CeFi victim behaviors~\cite{silicon}, prior hack or scam experiences do not seem to affect many DeFi users' security perceptions and confidence in DeFi services. Their strong financial motivations seem to outweigh security priorities and concerns. These observations also suggest that experience and education may be insufficient to help users employ better security practices (See \autoref{s:qual-rq3} and \autoref{s:quan-rq3}). Last, the primary concern for opposing DeFi regulation was paying tax. Considering that the primary objective of DeFi users is to maximize profit, this observation is unsurprising. Many expressed a willingness to sacrifice the potential security benefits of regulation (See \autoref{s:qual-rq4} and \autoref{s:quan-rq4}).

%% file: discuss.tex
\section{Discussion}
\label{s:discuss}

We present actionable recommendations based on our key findings, and discuss the limitations of the two studies.

\subsection{Recommendations}

\PP{Regulating DeFi to protect users} DeFi victims have a tendency to ignore security risks and pursue financial gains -- some victims experienced multiple scam or hack incidents as a result. Such behaviors diverge significantly from previous findings related to internet scams~\cite{chen2017securing, button2014not} where victims typically employ stronger protective measures after encountering frauds. Perhaps this difference can be explained partially by the analogy developed by Mills~\etal~\cite{mills2019preliminary}, Delfabbro~\etal~\cite{delfabbro2021cryptocurrency}, and Johnson~\etal~\cite{johnson2023cryptocurrency}: they present the idea that cryptocurrency users' trading practices and addictions often resemble gambling behaviors. 
Our results support their analogy. P13 mentioned ``\textit{Some of these DeFi projects are, in my opinion, basically [like] gambling}.'' If this analogy is accurate, then providing educational support and gaining more experience alone may not be effective in protecting DeFi users. Similar to how the internet gambling industry is heavily regulated to protect gamblers~\cite{miller2014regulation}, the DeFi industry may also need to mandate strong security controls through regulation and regular audit requirements.

Regulating DeFi is a complicated issue: 46.9\% of survey participants opposed the idea, expressing concerns about decentralization benefits being jeopardized through heavy regulation practices. Hence, the community must collectively work toward a decentralized form of regulation. For instance, the concept of a decentralized organization could be employed to facilitate autonomous management of DeFi project memberships~\cite{rethink-law}. Such organizations would conduct audits and enforce compliance through the means of verifying project team identity, analyzing whitepapers and roadmaps, and auditing project codes. Scam projects or projects that lack security controls would be rejected. Only those that pass the audits and compliance checks would be issued, e.g., a membership certificate -- users can check this information before safely engaging with a reliable DeFi service. 

\PP{Correcting security misconceptions}
We identified several security misconceptions. 
First, many users underestimate the importance of reviewing and revoking token approvals. Educational support and regular reminders are necessary to help users adopt best practices for re-configuring token grant limits to minimize risks.
MetaMask wallet, for example, offers comprehensive educational materials related to revoking token approvals~\cite{metamask-approval}.
In addition, we imagine a reminder feature that regularly prompts users to review their current approvals would also be effective.
Second, users often believe that DeFi platforms provide an equal level of security as the underlying blockchain technology -- such an inadequate security mental model needs to be improved, and users need to understand that DeFi platforms are just as prone to common online attacks.
Last, many users seem to believe that 2FA is consistently effective against various DeFi threats.
To help users understand the security protections offered through 2FA and its limitations, we suggest clarifying the covered threats in the 2FA settings of custodial wallets (e.g., Coinbase 2FA setting\footnote{https://accounts.coinbase.com/security/settings}) that users use to access DeFi services.

\subsection{Limitations}
This work has limitations because of its empirical nature.
Though applied validation questions about DeFi usage, we relied on self-reported data to recruit DeFi users, which could not prevent participants from giving desirable or repetitive survey responses.
In addition, participants' awareness and understanding of different DeFi hacks and scams may influence the self-reported type of incidents they suffered.
Another key limitation of our study is the absence of a standardized framework for analyzing the diverse range of DeFi services and applications, which may result in significant variations in user perceptions of their usefulness and security. This limitation restricts our ability to understand how specific DeFi users, such as borrowers or lenders, perceive these factors compared to others. Future studies need to consider these compound factors through the control of demographics.
Furthermore, to best accurately reflect the real-world distribution, we did not control the number of victim and non-victim participants. 
Therefore, we lacked sufficient responses from victims for more in-depth statistical analysis. We also note that we did not ask victims about their net profit or loss. Without this information, we cannot strongly claim that all victims are making short-sighted investment decisions -- some victims, despite being affected by several scams or hacks, may be accepting carefully calculated risks to continue investing in DeFi and generating net profit. Investigation of such a strategic group of victims would be an intriguing future work.
Lastly, ~\autoref{t:survey-demo} and~\autoref{a:yoe-asset} show 
our survey participants' yearly income and their asset distribution percentages in the crypto market and DeFi, respectively.
The risk-taking behavior revealed in this paper may not apply to users who are wealthy or arrange a significant percentage of income in DeFi.

%% file: relwk.tex
\section{Related Work} \label{s:relwk}

\PP{User studies in DeFi}
With the increasing popularity of DeFi, researchers have conducted various user studies within the DeFi ecosystems. For instance, Wang~\etal~\cite{wang:chi:2022} discovered that many DeFi users lack awareness of sandwich attacks and exhibit a high tolerance for them even after being educated about these attacks -- this observation is similar to our report about users' attitudes toward DeFi incidents. However, they also explain that DeFi traders believe they would learn ``how to avoid further losses'' from losing money, and by being attacked once, they will be motivated to learn and protect themselves next time. These perceptions, studied in the context of a single sandwich attack and non-victims, conflict with our analysis on real-world victims: despite being affected by various DeFi scams and hacks, many victims continued to use DeFi services without revising their security practices.
Additionally, Guan~\etal~\cite{guanexamining} identified misconceptions among DeFi users, such as the belief that stablecoin developers collaborate with regulatory entities like governments.
They also reported various perceived risks related to stablecoins. Our risk concern analysis, however, revealed that the aggregated stablecoin risk is one of the least concerning risks.

Feng~\etal~\cite{feng2023defi} explored users' perceptions of DeFi auditing and found that interview participants had difficulties interpreting technical audit reports, indicating a gap in the effectiveness of DeFi auditing. Chaliasos~\etal~\cite{chaliasos:icse:2024} conducted user studies with DeFi security practitioners to evaluate whether DeFi security tools meet their needs. Notably, these related works do not address our research questions, thereby motivating this paper.

\PP{Mitigations of DeFi incidents}
Several research approaches have been pursued to prevent DeFi hacks and scams before they occur. A significant direction includes identifying smart contract vulnerabilities through techniques such as fuzzing~\cite{he:ccs:2019, wustholz:fse:2020, nguyen:icse:2020, choi:icse:2021}, symbolic execution~\cite{luu:ccs:2016, krupp:sec:2018, manticore}, and static analysis~\cite{kalra:ndss:2018, tsankov:ccs:2018, so:oakland:2020}. These methods have successfully detected numerous smart contract vulnerabilities before they could be exploited.
Regarding efforts to mitigate scams, Cernera~\etal~\cite{cernera:sec:2023} discovered that 60\% of tokens on Ethereum and Binance Smart Chain last less than one day and highlighted patterns of rug-pulls. Huang~\etal~\cite{huang2023miracle} developed a prediction model to identify NFT rug-pull projects before incidents occur. Conversely, Wang~\etal~\cite{ERC-20R} focused on recovery from DeFi incidents. To revert exploit txs, they introduced new types of tokens and NFTs, named ERC-20R and ERC-721R, which allow for transaction reversal.

%% file: conclusion.tex
\section{Conclusion}
\label{s:conclusion}

Based on a semi-structured interview and a follow-up large-scale survey, we investigated DeFi users' security perceptions and the adequacy of commonly employed security practices.
Our results showed that DeFi users tend to have inadequate understanding of the security controls, and ignore security risks to pursue financial motivations. Many participants falsely believed 2FA (not supported in non-custodial wallets) can protect them from most of the common DeFi threats. Our investigation of victims revealed more worrying behaviors: many victims' security perceptions did not change after experiencing scams or hacks, and they continued to use other DeFi services to quickly recover their losses without revising security practices. The majority of such victims failed to take any remedy actions after the incident. Victim behaviors indicate that educational support and gaining more experience may have a limited impact -- a much stronger control, such as decentralized regulation, may be necessary to mandate adequate security practices and protect users.

%% file: append.tex
\appendix

\section{Interview Code Saturation Results}
\label{a:code-saturation-results}

\begin{figure}[!h]
    \centering
    \includegraphics[width=\columnwidth]{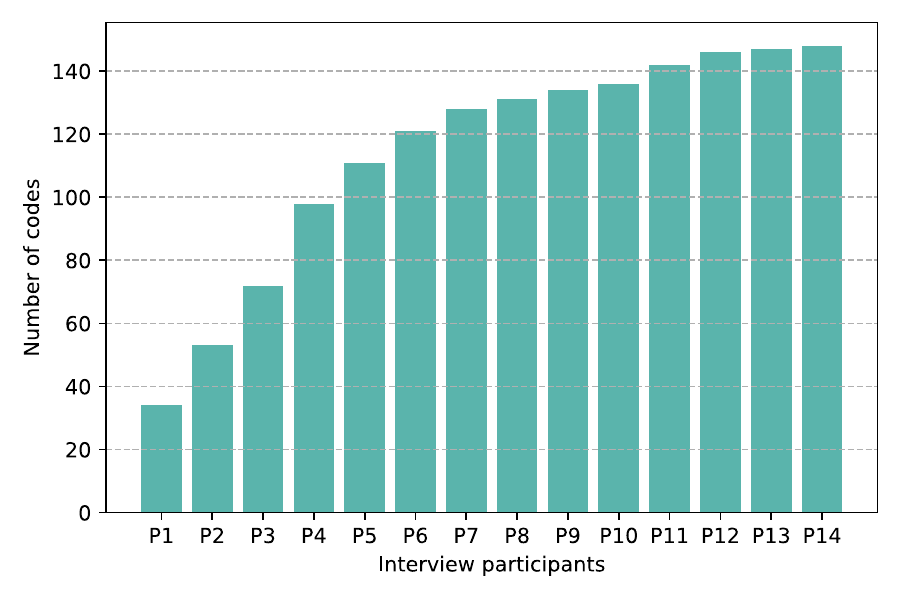}
    \caption{Code saturation status for the interview study}
    \label{fig:code}
\end{figure}

\section{Survey Results of Trading Experience and Asset Distribution}
\label{a:yoe-asset}

\begin{table}[!h]
    \begin{threeparttable}
        \centering
        \footnotesize
        \caption{Survey participants' YoE\tnote{1} in cryptocurrency and DeFi and their asset distribution.}
        \label{t:yoe-distribution}
        \begin{tabular}{ll|c|cc}
        \hlineB{3}
        \multirow{2}*{\textbf{Item}} & \multirow{2}*{\textbf{Property}} & {\textbf{All}} & {\textbf{Non-victim}} & {\textbf{Victim}}  \\
        \cline{3-5}
                                   &                   & $N$=493      & $N$=391   & $N$=102     \\
        \hlineB{3}
        \multirow{2}*{Crypto}      & Mean           & 4.6  & 4.5 & 5.1\\
        \multirow{2}*{YoE}         & Median         & 4.0  & 4.0 & 4.0\\
                                   & Std. deviation & 4.8 & 4.2 & 6.4\\
        \hline
        \multirow{2}*{\% of asset} & Mean           & 16.9  & 15.7 & 21.1\\
        \multirow{2}*{in Crypto}   & Median         & 10.0 & 10.0 & 15.0\\
                                   & Std. deviation & 17.9 & 17.1 & 20.2\\
        \hline
        \multirow{2}*{DeFi}        & Mean           & 2.3  & 2.2 & 2.5\\
        \multirow{2}*{YoE}         & Median         & 2.0  & 2.0 & 2.0\\
                                   & Std. deviation & 1.4  & 1.4 & 1.3\\
        \hline
        \multirow{2}*{\% of Crypto} & Mean          & 22.3  & 21.6 & 24.8\\
        \multirow{2}*{in DeFi}      & Median        & 10.0  & 10.0 & 13.5\\
                                    & Std. deviation& 28.5  & 28.6 & 28.1\\
        \hlineB{3}
        \end{tabular}
        \begin{tablenotes}
           \item [1] Abbreviated for \textit{Years of Experience}.
        \end{tablenotes}
    \end{threeparttable}
\end{table}

\clearpage

\onecolumn

\section{Statistical Results of DeFi Perceptions of Love/Pain Points}
\label{a:perception-test}

\begin{table*}[!ht]
\centering
\begin{singlespacing}
    \tabcolsep=6pt
    \footnotesize
    \caption{Results showing significant differences in Mann-Whitney U tests under Bonferroni correction for DeFi's importance distribution of \textit{love points}. The $p$-values and effect sizes in rank-biserial correlation $r$ are reported.}
    \label{t:love-test}
\begin{tabular}{|c|c|c|c|c|}
\hline
 & It's transparent & More opportunities & Active development & Good documentation \\
\cline{2-5} 
\textbf{Decentralization} & $p$ = 1.0e-5 & $p$ = 1.5e-5 & $p$ = 1.7e-4 & $p$ = 5.0e-4  \\
 & $r$ = -- 0.29 & $r$ = -- 0.27 & $r$ = -- 0.36 & $r$ = -- 0.33 \\
\hline
 & It's transparent & More opportunities & Active development \\
\cline{2-4} 
\textbf{It's secure} & $p$ = 6.9e-5 & $p$ = 1.0e-4 & $p$ = 3.8e-4 \\
 & $r$ = -- 0.29 & $r$ = -- 0.27 & $r$ = -- 0.36 \\
\cline{1-4} 
\end{tabular}
\end{singlespacing}
\end{table*}

\begin{table*}[!ht]
\centering
\begin{singlespacing}
    \tabcolsep=6pt
    \footnotesize
    \caption{Results showing significant differences in Mann-Whitney U tests under Bonferroni correction for DeFi's importance distribution of \textit{pain points}. The $p$-values and effect sizes in rank-biserial correlation $r$ are reported.}
    \label{t:pain-test}
\begin{tabular}{|c|c|c|c|c|c|c|}
\hline
 & Mobile using is poor & Lack of explanation & Lack of assets & Unclear gas allowance & No fiat gate & Bad UI \\
\cline{2-7} 
\textbf{Security in general} & $p$ = 4.3e-6 & $p$ = 3.1e-5 & $p$ = 5.4e-4 & $p$ = 3.5e-4 & $p$ = 1.4e-4 & $p$ = 5.5e-5 \\
 & $r$ = -- 0.36 & $r$ = -- 0.28 & $r$ = -- 0.23 & $r$ = -- 0.34 & $r$ = -- 0.28 & $r$ = -- 0.29 \\
\hline
 & Mobile using is poor  \\
\cline{2-2} 
\textbf{Transactions are costly} & $p$ = 1.6e-4 \\
 & $r$ = -- 0.29 \\
\cline{1-2} 
\end{tabular}
\end{singlespacing}
\end{table*}

\section{Statistical Results of DeFi Perceived Risks}
\label{a:risk-test}

\begin{table*}[!ht]
\centering
\begin{singlespacing}
    \tabcolsep=6pt
    \footnotesize
    \caption{Results showing significant differences in Mann-Whitney U tests under Bonferroni correction for DeFi's concern distribution of \textit{perceived risks}. The $p$-values and effect sizes in rank-biserial correlation $r$ are reported.}
    \label{t:risk-test}
\begin{tabular}{|c|c|c|c|c|c|c|}
\hline
 & Financial risks & Theft of private key & Phishing & Regulatory uncertainty & Instability of stablecoins & Airdrop scam \\
\cline{2-7} 
\textbf{Rug-pull} & $p$ = 1.1e-3 & $p$ = 3.7e-5 & $p$ = 1.6e-7 & $p$ = 1.6e-9 & $p$ = 1.5e-6 & $p$ = 4.7e-10 \\
 & $r$ = -- 0.14 & $r$ = -- 0.19 & $r$ = -- 0.25 & $r$ = -- 0.28 & $r$ = -- 0.24 & $r$ = -- 0.33 \\
\hline
 & Financial risks & Smart contract exploitation\\
\cline{2-3} 
\textbf{Airdrop scam} & $p$ = 4.8e-4 & $p$ = 5.3e-4 \\
 & $r$ = 0.19 & $r$ = 0.20 \\
\cline{1-3} 
\end{tabular}
\end{singlespacing}
\end{table*}

\clearpage

\twocolumn